\documentclass[aps,reprint,pra,floatfix,superscriptaddress,showpacs,lengthcheck,citeautoscript]{revtex4-1}
\usepackage{amsmath}
\usepackage{amssymb}
\usepackage{graphicx}
\usepackage{epstopdf}
\usepackage{bm}
\usepackage{color,soul}
\usepackage{times}
\usepackage{MnSymbol}

\newcommand{\ve}{\varepsilon}

\newcommand{\vf}{v_{\mathrm{F}}}

\newcommand{\bea}{\begin{eqnarray}}
\newcommand{\eea}{\end{eqnarray}}
\newcommand{\be}{\begin{equation}}
\newcommand{\ee}{\end{equation}}

\newcommand{\ci}{i}
\newcommand{\la}{\langle}
\newcommand{\ra}{\rangle}
\newcommand{\ket}[1]{| #1 \rangle}
\newcommand{\bra}[1]{\langle #1 |}

\renewcommand{\Im}{\mathrm{Im}}

\newcommand{\hw}{\hbar\Omega}

\begin{document}
\title{Hierarchy of Floquet gaps and edge states for driven honeycomb lattices}
\author{P. M. Perez-Piskunow}
\affiliation{Instituto de F\'{\i}sica Enrique Gaviola (CONICET) and FaMAF, Universidad Nacional de C\'ordoba, 5000 C\'ordoba, Argentina.}
\author{L. E. F. Foa Torres }
\affiliation{Instituto de F\'{\i}sica Enrique Gaviola (CONICET) and FaMAF, Universidad Nacional de C\'ordoba, 5000 C\'ordoba, Argentina.}
\author{Gonzalo Usaj}
\affiliation{Centro At{\'{o}}mico Bariloche and Instituto Balseiro,
Comisi\'on Nacional de Energ\'{\i}a At\'omica, 8400 S. C. de Bariloche, Argentina.}
\affiliation{Consejo Nacional de Investigaciones Cient\'{\i}ficas y T\'ecnicas (CONICET), Argentina.}
\begin{abstract}
%
%
Electromagnetic driving in a honeycomb lattice can induce gaps and topological edge states with a structure of increasing complexity as the frequency of the driving lowers. While the high-frequency case is the most simple to analyze we focus on the multiple photon processes allowed in the low-frequency regime to unveil the hierarchy of Floquet edge states. 
In the case of low intensities an analytical approach allows us to derive effective Hamiltonians and address the topological character of each gap in a constructive manner. At high intensities we obtain the net number of edge states, given by the winding number, with a numerical calculation of the Chern numbers of each Floquet band. 
Using these methods, we find a hierarchy that resembles that of a Russian nesting doll. This hierarchy classifies the gaps and the associated edge states in different orders according to the electron-photon coupling strength. For large driving intensities, we rely on the numerical calculation of the winding number, illustrated in a map of topological phase transitions.
The hierarchy unveiled with the low-energy effective Hamiltonians, along with the map of topological phase transitions discloses the complexity of the Floquet band structure in the low-frequency regime. The proposed method for obtaining the effective Hamiltonian can be easily adapted to other Dirac Hamiltonians of two-dimensional materials and even the surface of a three-dimensional topological insulator.
\end{abstract}
\date{\today}
 
\maketitle
%
%
\section{Introduction}
%
%
A topological material or system (e.g., a quantum Hall insulator or a topological insulator) has a bulk gap characterized by a topological invariant bearing a non-trivial value~\cite{Hasan2010,Bernevig2013}. The bulk-boundary correspondence establishes that when in contact with the vacuum (or a trivial material) the interface between the two media hosts conducting edge states~\cite{Hasan2010}. Interestingly, the number and chirality of the edge states are solely determined by the topological invariants computed for the \textit{bulk} systems. Recently, several studies signaled that topological edge states can be engineered in an ordinary material by applying a time-periodic driving~\cite{Oka2009,Kitagawa2010,Lindner2011}. This sparked the interest of diverse communities from graphene~\cite{Calvo2011,Zhou2011,Kitagawa2011,Iurov2012,SuarezMorell2012,Perez-Piskunow2014,Usaj2014} and related materials~\cite{Sie2015,Lopez2015}, to topological insulators~\cite{Dora2012,Calvo2015}, photonic crystals~\cite{Rechtsman2013}, and optical lattices~\cite{Goldman2014,Choudhury2014,Bukov2014,Bilitewski2014,Dasgupta2015,Qing2013,DAlessio2014,Goldman2015,Mori2015}, aiming to tackle a plethora of issues: characterization of these novel edge states~\cite{Perez-Piskunow2014,Usaj2014}, different signatures in magnetization and tunneling~\cite{Fregoso2014b,Dahlhaus2014}, the proper invariants entering the bulk-boundary correspondence~\cite{Rudner2013,Ho2014,Yang2014}, their statistical properties~\cite{Dehghani2014,Liu2014}, the role of interactions and dissipation~\cite{Seetharam2015,Iadecola2015,Dehghani2015} and the associated two-terminal~\cite{Gu2011,Kundu2014} and multiterminal (Hall) conductance ~\cite{Foa2014,Dehghani2015}.

Floquet theory~\cite{Sambe1973,Shirley1965,Grifoni1998,Platero2004,Kohler2005} is the prevalent tool for the study of time-periodic Hamiltonians. Within Floquet theory, the solutions of the time-dependent Schr\"odinger equation can be conveniently casted in terms of the solutions of an eigenvalue problem in a higher-dimensional space, the so-called Floquet space~\cite{Kohler2005,Sambe1973} which is the direct product between the usual Hilbert space and the space of time-periodic functions with period $T=2\pi/\Omega$. The increased dimensionality is at the heart of the richness arising in the Floquet quasienergy spectra. Notably, when the driving opens a gap between two adjacent Floquet replicas, other replicas (associated to different number of photons) develop a hierarchy of ever smaller gaps, each of them hosting chiral edge states. The ensuing structure, which reminds us of Russian nesting dolls, progressively unfolds as higher-order inelastic processes are explored.

While for high-frequency driving, i.e., of the order of or larger than the bandwidth, the system's stroboscopic evolution~\cite{Kitagawa2011} can be elegantly described by an effective time-independent Hamiltonian~\cite{Goldman2014,Goldman2015,Eckardt2015}, the opposite low-frequency regime is trickier to deal with, but might be experimentally more feasible for many materials like three-dimensional topological insulators~\cite{Wang2013}, graphene~\cite{Calvo2011,Perez-Piskunow2014, Usaj2014}, or other two-dimensional materials~\cite{Sie2015}. Moreover, it is in this regime that the mentioned nesting structure appears and the determination of an effective Hamiltonian and the characterization of the associated chiral edge states becomes more challenging.

Here we address the nesting structure of the bulk gaps and associated edge states in the Floquet quasienergy spectra of honeycomb lattices. To do this we rely on the fact that these gaps follow a hierarchy in which the gaps' widths depend on the order of the inelastic processes originating them. This allows us to determine the number of edge states by looking  first at the largest energy scale (largest gap) and progressively moving into the smaller (higher-order) gaps towards the gap center. The hierarchy unfolds as new edge states bridge the smaller gaps. Honeycomb lattices illuminated by an intense circularly polarized laser have attracted much attention in this context~\cite{TenenbaumKatan2013,Gomez-Leon2014} but a detailed analysis for frequencies spanning both high- and low-frequency regimes is missing. Here we provide a systematic derivation of the effective Hamiltonians at the crossings between Floquet bands together with analytical expressions for the associated contributions to the Chern numbers.

This work is organized as follows. In Sec.~\ref{sec:floquet} we present the Floquet Hamiltonian for an irradiated honeycomb lattice. In Sec.~\ref{sec:chern} we discuss the calculation of the Chern numbers of the Floquet bands in terms of the low-energy (Dirac) Hamiltonian and explain the hierarchy of the corresponding edge states. The case of large driving intensity and frequency is analyzed in Sec.~\ref{sec:map}, where a full map of the Chern number is obtained by a direct numerical calculation using the bulk tight-binding Hamiltonian. This enables us to show a phase diagram of the topological phase transitions for a wide range of frequencies and intensities of the driving field.
%
%
\section{\label{sec:floquet}Driven honeycomb lattice}
%
Let us consider a general system with a Hamiltonian $\mathcal{H}_0$ (time-independent) in the presence of a time-periodic perturbation $\mathcal{V}(t)$. The full Hamiltonian $\mathcal{H}(t)=\mathcal{H}_0+\mathcal{V}(t)$ satisfies $\mathcal{H}(t+T)=\mathcal{H}(t)$, where the period $T=2\pi/\Omega$ is determined by the driving frequency $\Omega$. Floquet's theorem guarantees the existence of a set of solutions of the time-dependent Schr\"odinger equation of the form $\ket{\psi_{\alpha}(t)}=\exp(-\ci\varepsilon_{\alpha}t/\hbar)\ket{\phi_{\alpha}(t)}$, where $\ket{\phi_{\alpha}(t)}$ has the same time periodicity as the Hamiltonian, $\ket{\phi_{\alpha}(t+T)}=\ket{\phi_\alpha(t)}$~\cite{Shirley1965,Grifoni1998,Platero2004,Kohler2005}---this is the equivalent of the usual Bloch theorem for systems that are periodic in real space.
The Floquet states $\ket{\phi_{\alpha}(t)}$ are the solutions of the eigenvalue equation $\mathcal{H}_F\ket{\phi_{\alpha}(t)}=\varepsilon_{\alpha}\ket{\phi_{\alpha}(t)}$, where $\mathcal{H}_F=\mathcal{H}-\ci\hbar \frac{\partial}{\partial t}$ is the so-called Floquet Hamiltonian and $\varepsilon_{\alpha}$ is the Floquet quasienergy.

It is customary, and useful, to introduce the notion of the Floquet space, formed by the direct product between the Hilbert space and the space of time-periodic functions with period $T$ (spanned by the functions $e^{\ci n\Omega t}$ with $n=0,\pm1,\pm2,\dots$), so that $\ket{\phi_{\alpha}(t)}=\sum_n e^{\ci n\Omega t} \ket{u^\alpha_n}$.
When written in this basis, the Floquet Hamiltonian $\mathcal{H}_F $ is a time-independent infinite matrix $\mathcal{H}_F^\infty$ with copies of $\mathcal{H}_0$ in the diagonal blocks or Floquet replicas (fixed $n$). Each diagonal block is shifted in energy by $n\hbar\Omega$. The time-dependent perturbation enters only (if it has zero time-averaged value) in the off-diagonal blocks that couple the different Floquet replicas.

In analogy with the concept of the Brillouin zone for Bloch electrons, the quasienergies can be restricted to a Floquet zone. Indeed, for every solution $\ket{\phi_{\alpha}(t)}$ with quasienergy $\varepsilon_\alpha$ one can construct another solution $\ket{\phi_{\alpha m}(t)}=\exp(-\ci m\Omega t)\ket{\phi_{\alpha}(t)}$ with quasienergy $\varepsilon_{\alpha m}=\varepsilon_\alpha+m\hbar\Omega$, that corresponds to the same physical state $\ket{\psi_{\alpha}(t)}$. Therefore, the eigenvalues are repeated at intervals of $\hbar\Omega$ and they could be restricted to the interval $(-\hbar\Omega/2,\hbar\Omega/2]$. 
While this reduced zone scheme is the usual choice, we find it more convenient and more insightful, for reasons that will become clear below, to work in the extended zone scheme. 
In that case, to better interpret the results, a useful magnitude that complements the spectral information (see below) is the time-averaged ``local'' density of states which can be computed as the density of states associated to the Floquet Hamiltonian projected on the $n=0$ Floquet subspace~\cite{Oka2009,Zhou2011}
%
\be
\bar{\rho}_a(\ve)=\sum_\alpha \delta(\ve-\ve_\alpha) |\langle a|u^\alpha_0\rangle|^2\,,
\label{density}
\ee
%
where $\ket{a}$ is an arbitrary state of the Hilbert space. In the sum, the full set of quasienergies $\ve_\alpha$ is kept to ensure that for vanishing intensity of the time-periodic potential (and hence of the coupling between the Floquet replicas) the original density of states of the unperturbed system is recovered. Equation~\eqref{density} can also be casted in terms of the Floquet-Green function~\cite{Calvo2013,Usaj2014}. It is worth noting that recent works point out the key role played by the time averaged component of the Floquet eigenstates~\cite{Usaj2014,Foa2014,Fregoso2014b,Kundu2014}, particularly when analyzing the transport response of the driven system~\cite{Foa2014}.
%
%
\subsection{\label{sub:floquet}Floquet-Bloch Hamiltonian}
%
A honeycomb lattice with a single orbital per site can be described by the following \emph{tight-binding} Hamiltonian
%
\begin{equation}
{\cal H}_\mathrm{tb}(t)=\sum_{i}\epsilon_{i}^{{}}\,c_{i}^{\dagger}c_{i}^{{}}-\sum_{\left\langle i,j\right\rangle }[\gamma_{ij}(t)\,c_{i}^{\dagger}c_{j}^{{}}+\mathrm{h.c.}]\,.
\end{equation}
%
Here $c_{i}^{\dagger}$ and $c_{i}^{{}}$ are the electronic creation and annihilation operators at site $i$ with energy $\epsilon_{i}$, respectively, and $\gamma_{ij}$ is the nearest-neighbors hopping matrix element. We neglect the spin degree of freedom throughout this work as it does not play any role.

The effect of the circularly polarized electromagnetic field $\bm{E}(t)$ can be described in a gauge such that $\bm{E}(t)=-(1/c)\, \partial \bm{A}/ \partial t$, where $\bm{A}(t)=A_0 (\cos\Omega t,\sin\Omega t)$ is the vector potential---this describes the situation of normal incidence. Hence, the time-dependent field enters the Hamiltonian through the hopping matrix elements (Peierls substitution):
%
\begin{equation} 
  \gamma_{ij}(t)= \gamma\exp\left(\ci\frac{2\pi}{\Phi_0}\int_{
  \bm{r}_i}^{\bm{r}_j}\bm{A}(t)\cdot\mathrm{d}\bm{\ell}\right)\,,
  \label{gama}
\end{equation}
%
where $\Phi_0$ is the magnetic flux quantum.

Following a similar procedure as in Refs.~\cite{Koghee2012,Delplace2013} we arrive at the Floquet-Bloch Hamiltonian, $\mathcal{H}_{F}(\bm{k})=\sum_{m,n} H_{m,n} + \delta_{m,n} \hw\,I$, where $H_{m,n}=1/T\int_0^\infty e^{i \Omega t (n-m)} H(t) \mathrm{d}t$ is the $(n-m)$ Fourier component of the time-dependent Hamiltonian. Each diagonal block has copies of $\mathcal{H}_0$ that account for the Floquet replicas; the hoppings between different lattice sites within the same replica are the zeroth Fourier components of $\gamma_{i,j}(t)$. This is proportional to $\gamma J_0(z)$ up to a phase that depends on the direction of the hopping, where $J_0(x)$ is the zeroth order Bessel function, $z=A_0 a_c 2\pi/\Phi_0 $ denotes the field intensity from now on, and $a_c$ is the distance between nearest neighbors in the honeycomb lattice. This dependence on $J_0(z)$ will lead to many interesting behaviors of the topological characteristics (of any driven lattice) when the intensity reaches a root of $J_0(x)$. The first root at $z_{0,1}\simeq2.4048$ leads to a topological phase transition that is further explained in Sec.~\ref{sec:map}.
%
%
\subsection{Low-energy Hamiltonian}
%
Close to the Dirac points ($K$ and $K'$ points), the band structure of the honeycomb lattice is well described by a Dirac Hamiltonian, 
%
\begin{equation}
  \begin{array}{ll}
    \mathcal{H}(t)=&\hbar \vf\left[ \sigma_x \left(k_x\!+\!\frac{e}{\hbar c}A_0\cos\Omega t\right)\right.\\
    &+\left.s\,\sigma_y \left(k_y\!+\!\frac{e}{\hbar c}A_0\sin\Omega t\right) \right]\,,
  \end{array} \label{eq_laser_graphene_Hamiltonian}
\end{equation}
%
where $\vf$ denotes the Fermi velocity, $\bm{\sigma}=(\sigma_x,\sigma_y)$ are the Pauli matrices for the pseudospin degree of freedom, and $s=\pm1$ is the valley index.

For the $K$ valley ($s=1$) we obtain the Floquet Hamiltonian 
%
\begin{equation}
\begin{array}{l}
  \mathcal{H}^\infty_{F}(\bm{k})= 
  \left(
  \begin{array}{cccccc}
  \ddots & \vdots & \vdots & \vdots & \vdots & \udots \\
  \cdots & \hbar\Omega & \hbar \vf k_- & 0 & 0 & \cdots\\
  \cdots & \hbar \vf k_+ & \hbar\Omega& \frac{e \vf}{c}A_0 & 0 & \cdots\\
  \cdots & 0 & \frac{e \vf}{c}A_0 & 0 & \hbar \vf k_- & \cdots\\
  \cdots & 0 & 0& \hbar \vf k_+ & 0 & \cdots\\
  \udots & \vdots & \vdots & \vdots & \vdots & \ddots
  \end{array}
  \right)\,,
\end{array} \label{eq:Floquet_Ham_k}
\end{equation}
%
with $k_{\pm}=k_x\pm \ci k_y$. Since the external driving is harmonic, and in this approximation it enters linearly in the Hamiltonian, only the Floquet replicas differing in $\pm1$ photon will be directly coupled with a relative strength $\eta=e\vf A_0/c\hw $ [in connection with the lattice Hamiltonian $\eta=(3\gamma/2\hw)z$]. Higher-order couplings between two replicas, $m$ and $n$ are indirect and of order $\mathcal{O}(\eta^{|n-m|})$.
%
%
\section{\label{sec:chern}Hierarchy of driving induced gaps and edge states} 
%
\begin{figure}[!tbp] 
  \includegraphics[width=\columnwidth]{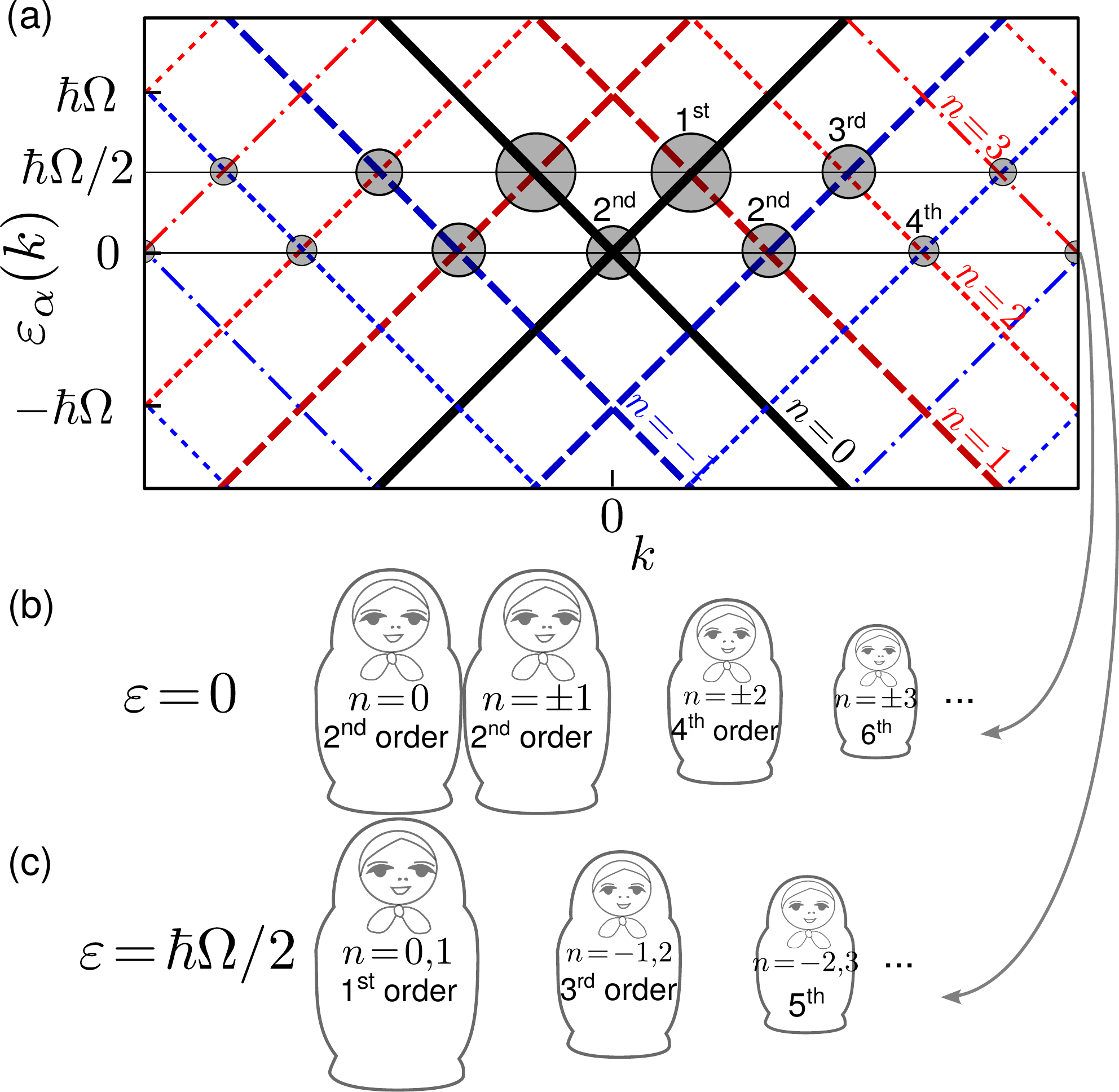}
  \caption{\label{fig:sketch}(a) Sketch of the dispersion of the first replicas around $n=0$. The crossings occur at the Floquet zone center, $\ve=0$, and at the Floquet zone borders $\ve=\pm\hbar\Omega/2$ are depicted with circles. Note that for $\ve=0$ the crossings involve replicas where $m+n=0$ and are of order $\eta^{|n-m|}$ ($|n-m|$ even), while in the case $\ve=\hw/2$ the crossings involve replicas where $m+n=\pm1$ and are of the order $\eta^{|n-m|}$ ($|n-m|$ odd). (b) and (c) Cartoon representations of different crossings for $\ve=0$ and $\hw/2$ respectively, ordered hierarchically according to their magnitude. The special case of the first doll in (b) represents the anticrossings at the Dirac point of the $n=0$ replica. This occurs because of a second-order process involving the emission and reabsorption of a photon.}
\end{figure}
%
%
The Floquet theory outlined in the previous section enables a simple picture of how the driving (in our case circularly polarized light) can lead to laser-induced gaps~\cite{Oka2009,Calvo2011,Zhou2011,Savelev2011}. Here we briefly highlight a few points that will be useful later on. 
We start considering the low-energy Hamiltonian of Sec.~\ref{sec:floquet}-A. For vanishing driving strength, we have the Floquet spectra represented in Fig.~\ref{fig:sketch}(a) (we take here a projection along a particular $k$ direction around the $K$ point). The effects of the external driving are expected to be important wherever the Floquet replicas corresponding to different values of the Fourier index $n$ become degenerate. This happens at half-integer multiples of $\hbar\Omega/2$. In Fig.~\ref{fig:sketch}(a) the crossings at $\ve_0=0$ and $\ve_{1/2}=\hbar\Omega/2$ are marked with gray circles. Interestingly, for circularly polarized light all these degeneracies are lifted (including the degeneracy between the bands with $n=0$ at $\ve=0$) with different strengths. In the low intensity limit ($\eta\ll1$), the magnitude of each anticrossing [of order $\mathcal{O}(\eta^{\Delta n})$] is ruled by the difference $\Delta n$ among the associated replicas, thereby establishing a hierarchy. This is schematically represented in Fig.~\ref{fig:sketch} (b) and (c).

Once the degeneracies develop into gaps, something interesting in the physics of topological systems happens: Edge states develop within each anticrossing and these states can co-exist with the continuum spectrum provided by other Floquet bands (these bands also have a gap of smaller width). The chirality and the robustness to disorder of such Floquet edge states were explicitly shown in Ref.~[\onlinecite{Perez-Piskunow2014}] and more recently, other authors pointed out that this could be a general fact also in time-independent systems \cite{Baum2015}. In the following we will exploit the structure shown in Fig.~\ref{fig:sketch} to systematically and progressively unfold our Floquet Russian nesting doll. At each step we will obtain an effective Hamiltonian describing the corresponding anticrossing, and the number and chirality of the edge states bridging it. The latter requires the determination of the relevant topological invariants that we briefly discuss in the next subsection. We then follow with our results for the low- and high-frequency regimes.
%
%
\subsection{Topological invariants for Floquet bands}
%
The Chern number associated to a given Floquet-Bloch band $\alpha$ is given by
%
\begin{equation}
  \begin{array}{lll}    
  C_\alpha&= \dfrac{\ci}{\strut 2\pi} \displaystyle \oint_\mathcal{C}\, \bra{u_{\alpha\bm{k}}}\bm{\nabla}_{\bm{k}}\ket{u_{\alpha\bm{k}}}\cdot d\bm{k}\\
  &=\dfrac{1}{\pi}  \Im\displaystyle\int_{\mathrm{BZ}}\,  \la\partial_{k_y}u_{\alpha\bm{k}}|\partial_{k_x}u_{\alpha\bm{k}}\ra\,d^2k \,,
  \end{array} \label{full_chern_0}
\end{equation}
%
where $\ket{u_{\alpha\bm{k}}}$ is the periodic part of the Bloch eigenfunction and $\mathcal{C}$ is the contour of the Brillouin zone (BZ)~\cite{Thouless1982}. Alternatively, Eq.~\eqref{full_chern_0} can be cast in the form
%
\be 
C_\alpha=\frac{1}{2\pi}\int_{\mathrm{BZ}}\,  \bm{\Gamma}_{\alpha\bm{k}}\cdot d\bm{S}_{\bm{k}} \,,
\label{full_chern_1}
\ee
%
with
%
\be
\bm{\Gamma}_{\alpha\bm{k}}=\Im\sum_{\beta\ne \alpha}\frac{\bra{u_{\alpha\bm{k}}}\bm{\nabla}_{\bm{k}}H_{\bm{k}}\ket{u_{\beta\bm{k}}}\times\bra{u_{\beta\bm{k}}}\bm{\nabla}_{\bm{k}}H_{\bm{k}}\ket{u_{\alpha\bm{k}}}}{(\ve_{\alpha\bm{k}}-\ve_{\beta\bm{k}})^2}\,,
 \label{full_chern_2}
\ee
%
where $\Gamma_{\alpha\bm{k}}$ is the Berry curvature. The peaks in the Berry curvature that occur at the points in the BZ where the bands are quasi degenerate yield the main contribution to $C_\alpha$. If the curvature decays fast enough (which happens when $\eta\rightarrow0$) the sum of these contributions is the exact calculation of $C_\alpha$. We will make use of this fact in Sec.~\ref{sec:low_energy} where an effective Hamiltonian is derived for the quasi degenerate subspace. We also note that though the topological invariants may seem very abstract objects they have recently been measured in cold matter experiments~\cite{Aidelsburger2015}.

In periodically driven systems to accurately account for the edge states one must rely on the winding number $W(\varepsilon)$ due to the infinite periodicity of the Floquet spectrum~\cite{Kitagawa2010,Rudner2013}. When the winding number is evaluated, inside a gap counts for the net number of chiral edge states. In connection with the Chern number, the difference of winding numbers evaluated at energies enclosing a band, yields the Chern number of that single band. From now on, we will only need to evaluate the winding number in the two distinct Floquet band gaps, the gap at the center of the Floquet zone [$W(\ve_0)=W(0)$], and the gap at the edge of the Floquet zone [$W(\ve_{1/2})=W(\hw/2)$].
This topological invariant can be obtained in terms of the evolution operator but, here we use an alternative approach proposed in Ref.~\cite{Rudner2013}, that consists in truncating the Floquet Hamiltonian between the replicas $-M$ and $M$ up to a sufficiently large $M$ (note that each extra replica adds two bands to the Floquet spectrum). The difference between the number of chiral edges states between the $\alpha$ and the $(\alpha+1)$ Floquet bands will be given by 
\begin{equation}
  W(\ve_\alpha)=\sum_{\beta=-(2M+1)}^{\alpha} C_\beta\,,
\end{equation}
for a quasienergy $\ve_\alpha$ inside the gap, provided that enough Floquet replicas are counted until the sum converges. This happens when taking a larger $M$ leaves $W(\ve_\alpha)$ unchanged, meaning that all relevant crossings between different replicas are included in the Floquet zone. Notice then that the continuum Dirac model is only appropriated as an approximation and requires  a finite number of replicas.

A direct evaluation of Eqs.~\eqref{full_chern_1} and~\eqref{full_chern_2} usually requires the use of numerics and the highly peaked Berry curvature renders the calculation easier for high frequencies. In this regime we can characterize the topological properties of the Floquet bands and the corresponding edge states using the bulk Floquet Hamiltonian, as seen in Sec.~\ref{sec:map}.

\subsection{\label{sec:low_energy}Multiple photon processes for low-frequency driving}
%
In this section we will apply a consistent method to obtain the number of edge states inside the driving induced gaps for the particular case of the honeycomb lattice. To do this we will take advantage of the hierarchy of these gaps, which scale as a power of $\eta$ with the exponent being the number of photon processes.

To obtain the winding numbers $W(\ve_0)$ and $W(\ve_{1/2})$ associated to the driving induced gaps at the Floquet zone center and at the Floquet zone edge, respectively, we must calculate the Chern numbers of all the Floquet bands below them. As outlined by Eq.~\eqref{full_chern_2} the main contributions to the Chern number of each band comes from the points in the $\bm{k}$~space where the energies are nearly degenerate. For a vanishing intensity the degeneracies will appear at the crossings of the Floquet replicas. When turning the electromagnetic field on, all the degeneracies will be lifted, opening gaps at every avoided crossing.

Let us use the limit of vanishing intensity to calculate the Chern number $C_\alpha$ of the $\alpha$ band. This can be obtained as the sum of all the contributions from the $\bm{k}$-space regions where an avoided crossing occurs. We will denote the contribution coming from a point $\bm{k}_{p,\alpha}$ where the $\alpha$ band has avoided crossings with the $(\alpha+1)$ band as $c^{\mathrm{up}}_{p,\alpha}$, and if an avoided crossing occurs at a (possibly different) point $\bm{k}_{p',\alpha}$ with the $(\alpha-1)$ band it will be denoted by $c^{\mathrm{low}}_{p',\alpha}$. So the sum that yields the Chern number is $C_\alpha=\sum_{p} c^{\mathrm{up}}_{p,\alpha}+\sum_{p'} c^{\mathrm{low}}_{p',\alpha}$. Here each contribution is obtained from Eq.~\eqref{full_chern_1} integrating only near the avoided crossing --for any finite intensity this is an approximate result but taking the limit where the intensity goes to zero the calculation becomes exact.

Since each avoided crossing means a contribution to the Chern number for the bands above and below it with opposite signs ($c^{\mathrm{low}}_{p+1\alpha}=-c^{\mathrm{up}}_{p,\alpha}$), when adding all the Chern numbers up to the $\alpha$ band to obtain the winding number, most of these contributions will cancel out except for the lasts ones (note that the first band in the truncated Floquet spectrum has no band below and no crossings $c^{\mathrm{low}}_{p,\alpha}$; see Fig.~\ref{fig:edge}).
So we obtain $W_\alpha=\sum_p c^{\mathrm{up}}_{p,\alpha}$. Since these contributions are the only ones that determine the number and chirality of the edge states we can drop the superscript in the following.

We can see in Fig.~\ref{fig:sketch} that the degeneracies appear at $k_{p,0}=2p\,k_0$ for the gap at the Floquet zone center ($\ve_0$), and $k_{p,1/2}=(2p+1)k_0$ for the gap at the Floquet zone edge ($\ve_{1/2}$), being $\vf k_0=\Omega/2$ and $p$ being an integer number.
In order to get the contribution near the anti-crossing between two replicas it is sufficient to derive a $2\times2$ effective Hamiltonian, valid close to $k_{p,\beta}$, with $\beta$ either zero or one-half.
By writing this Hamiltonian as
%
\be
\mathcal{H}^{\mathrm{eff}}_F(\bm{k},p,\beta)=\vf \bm{h}_{p,\beta}(\bm{k})\cdot\bm{\sigma}+\ve_\beta \bm{I}\,,
\ee
%
one can obtain the contribution to the Chern number by calculating
%
\be 
c_{p\beta}=\frac{1}{4\pi}\int\, \hat{\bm{h}}_{p,\beta}\cdot\left(\partial_{k_x}\hat{\bm{h}}_{p,\beta}\times\partial_{k_y}\hat{\bm{h}}_{p,\beta}\right)\, d^2k\,.
\label{cp}
\ee
%
with $\hat{\bm{h}}_{p,\beta}=\bm{h}_{p,\beta}/|\bm{h}_{p,\beta}|$.

To obtain an explicit form for $\mathcal{H}^{\mathrm{eff}}_F(\bm{k},p,\beta)$ we start by making a unitary transformation of the pseudospin basis. The basis $\left\{1/\sqrt{2},\pm\exp(\ci\theta_{\bm{k}})/\sqrt{2}\right\}^\mathrm{T}$ diagonalizes every diagonal block in Eq.~\eqref{eq:Floquet_Ham_k} (Floquet replica) describing the Dirac cone with eigenvalues $\pm\hbar\vf|k|$ shifted by $n\hbar\Omega$ for the $n\text{-th}$ Floquet replica. As depicted in Fig.~\ref{fig:sketch} replicas indexed by $m$ and $n$ will cross at $\ve_0$ when $m+n=0$, while the crossing will occur at $\ve_{1/2}$ if $m+n=1$. Hence, we must calculate the effective coupling between the replicas ($-m$) and ($m$) or ($m+1$) according to whether we are evaluating $c_{m,0}$ or $c_{m,1/2}$. This is achieved by 
a standard procedure based on the projected Green's function (or decimation procedure).
Namely, if $G_F(\omega,\bm{k})$ denotes the Floquet Green's function, $G_F(\omega,\bm{k})=[\omega\bm{I}-\mathcal{H}^\infty_F(\bm{k})]^{-1}$, we define the effective Hamiltonian, in this case, as
%
\be
\mathcal{H}^{\mathrm{eff}}_F(\bm{k},p,\beta)=\mathcal{H}^0_F(\bm{k},p,\beta)-\tilde{G}_F^{-1}(\beta\hbar\Omega,\bm{k}_{p\beta})\,,
\ee
%
where $\tilde{G}_F(\omega,\bm{k}_{p,\beta})=P_{p,\beta}^\dagger G_F(\omega,\bm{k}_{p,\beta}) P_{p,\beta}$ is the Green's function projected on the degenerate subspace ($n+m=0$ or $m+n=1$) and evaluated at the crossing $\bm{k}_{p,\beta}$, $P_{p,\beta}$ is the corresponding projector operator, and $\mathcal{H}^0_F(\bm{k},p,\beta)$ is the projected Floquet Hamiltonian in the absence of the radiation field. 
Excluding the special case of the crossing at $\bm{k}_{0,0}$ treated later, one readily finds, to the lowest non-trivial order in $\eta$ (see note~\footnote{\label{note}To keep the expressions simple we have neglected corrections of order $\eta^2$ for the $z$ components of $\bm{h}$. These corrections come from a renormalization of the Floquet replicas and do not modify the contribution to the Chern number.}), that
%
\begin{eqnarray}
\nonumber
   \bm{h}_{p\beta}(\bm{k})&=&\eta^{s_p^\beta} a_{p}^\beta k_0 \left[\cos{(s_p^\beta\theta_{\bm{k}}})\, \hat{\bm{x}}+\sin{(s_p^\beta\theta_{\bm{k}}})\,\hat{\bm{y}}\right]\\
&&+(-k+s_p^\beta k_{0})\hat{\bm{z}}
\label{little_h}
\end{eqnarray}
%
where $s_p^0=2p$, $s_p^{1/2}=2p+1$, $\tan\theta_{\bm{k}}=k_y/k_x$ and $a_p^\beta$ is a numerical factor (see the Appendix). In order to calculate the contribution to the Chern number, Eq.~\eqref{cp}, it is necessary to transform back to a $\bm{k}$-independent basis since the unitary transformation we used depends on $\theta_{\bm{k}}$. This implies a rotation of the effective field $\tilde{\bm{h}}_{p\beta}=\bm{R}(\theta_{\bm{k}})\bm{h}_{p\beta}$.
Using polar coordinates we have, 
%
\begin{eqnarray}
\nonumber
c_{p\beta}&=&\frac{1}{4\pi}\int \tilde{\bm{h}}_{p\beta}\cdot\left(\partial_\theta \tilde{\bm{h}}_{p\beta}\times \partial_k \tilde{\bm{h}}_{p,\beta}\right)\frac{1}{|\tilde{\bm{h}}_{p,\beta}|^{3}}\, dk\,d\theta\\
\nonumber
&=&\frac{1}{4\pi}\int \bm{h}_{p,\beta}\cdot\left(\partial_\theta \bm{h}_{p\beta}\times \partial_k \bm{h}_{p,\beta}\right)\frac{1}{|\bm{h}_{p,\beta}|^{3}}\, dk\,d\theta\\
\nonumber
&&+\frac{1}{4\pi}\!\int\! \bm{h}_{p,\beta}\cdot\left(\bm{R}^{-1}\partial_\theta\bm{R}\, \bm{h}_{p,\beta}\times \partial_k \bm{h}_{p,\beta}\right)\frac{1}{|\bm{h}_{p,\beta}|^{3}}\, dk\,d\theta\,.\\
\label{c_rotado}
\end{eqnarray}
%
Notice that we took advantage of the fast convergence of the integrands and extended the integration to the entire $\bm{k}$ space. 
The last integral gives zero for $\bm{h}_{p,\beta}$ of the form of Eq.~\eqref{little_h}, while the other can be done explicitly to obtain
%
\begin{equation}
c_{p\beta}=\frac{s_p^\beta}{2} \left(1+\frac{1}{\sqrt{\left(\eta^{s_p^\beta} a_{p}^\beta/s_p^\beta\right)^2+1}}\right)\,.
\end{equation}
%
Retaining the lowest order in $\eta$ consistent with the approximation made to obtain $\bm{h}_{p,\beta}$, we get
%
\begin{equation}
c_{p\beta}=s_p^\beta\,.
\label{eq:identity}
\end{equation}
%
This is one of our central results. The same derivation can be obtained for the expansion around the $K'$ valley, and the total contribution (up to the proper order) to $C_\alpha$ is twice $c_{p,\beta}$, one per each valley. The equality \eqref{eq:identity} could have been anticipated from Eq.~\eqref{little_h} if one recalls that $c_{p,\beta}$ is related to the number of times $\bm{h}_{p,\beta}(\bm{k})$ winds around the Bloch sphere as $\bm{k}$ explores the Brillouin zone. The angular dependence of $\bm{h}_{p,\beta}(\bm{k})$ is related to the effective coupling between the two degenerate replicas through the intermediate ones. From the decimation procedure one can infer that the factor in the angular dependence equals the number of replicas decimated plus one or, in other words, it is the difference between the Floquet indices of the two replicas involved in the avoided crossing. The latter makes clear that $|s_p^\beta|$ is the order of the photon processes that lead to the avoided crossing . Following this algorithm when looking at the next crossing, $p+1$, the involved replicas will be $+2$ replicas apart, so $s_p^{\beta+1}=s_p^\beta+2$.
%
\begin{figure*}[!tbp]
  \includegraphics[width=\textwidth]{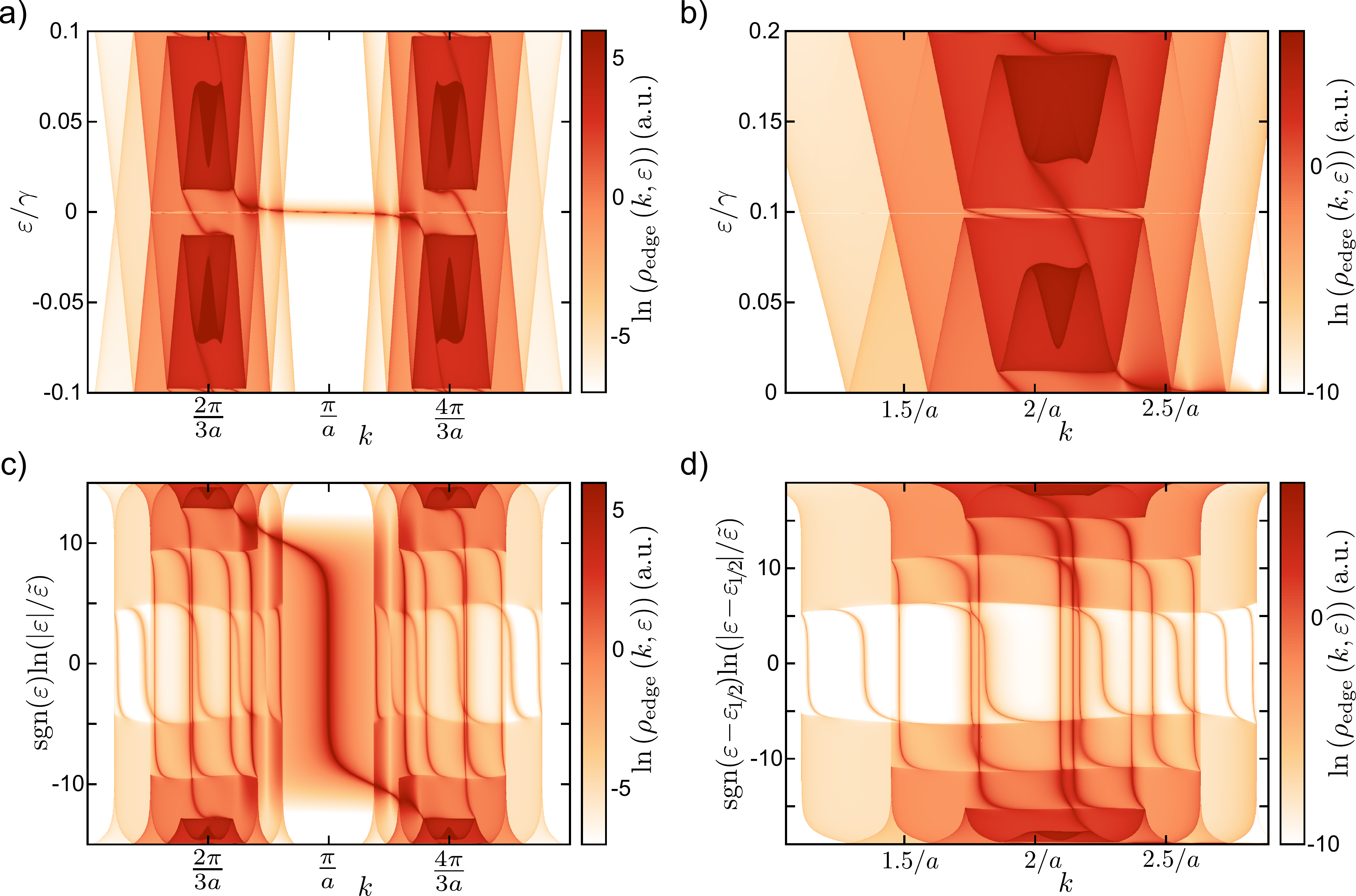}
  \caption{\label{fig:loge} (Color online) $k$-resolved local density of states near the edge of a semi-infinite honeycomb-lattice plane (in logarithmic scale in color). The plane is irradiated with $\hbar\Omega=0.2\gamma$ and $z=0.05$. In (a) the gap at $\ve_0$ is shown for both the $K$ and $K'$ valleys; in (b) the gap at $\ve_{1/2}$ is shown near the $K$ valley only. In the lower panel (c) the energy scale is expanded exponentially around $\ve_{0}=0\gamma$ up to a minimum cutoff energy $\tilde{\ve}\simeq3\times10^{-8}\gamma$. In panel (d) the energy is expanded exponentially around $\ve_{1/2}=0.1\gamma$ up to a minimum cutoff energy $\tilde{\ve}\simeq6\times10^{-10}\gamma$ [meaning that the interval $(-\tilde{\ve},\tilde{\ve})$ is not shown]. The lower panels (c) and (d) show the nested hierarchy in powers of $\eta$ of the developed gaps and their edge states.}
\end{figure*}
%

The only exception to this rule is the particular case of $c_{0,0}$ which only comes from the renormalization of the $m=0$ replica and there are no intermediate replicas involved. In this case we have 
%
\begin{equation}\label{eq:h_00}
\bm{h}_{0,0}(\bm{k})=- 2\eta^2 k_0\, \hat{\bm{x}}+ k\, \hat{\bm{z}}\,.
\end{equation}
It is clear from the above expression that the value of $c_{0,0}$ is determined by the last integral in Eq.~\eqref{c_rotado}, leading to
%
\begin{equation}
c_{0,0}=-\frac{1}{2}\,.
\end{equation}
%
Since we must count both Dirac cones ($K$ and $K'$ valleys) to get the total contribution to the Chern number, we get a total of $-1$ for the edge state connecting the $K$ and $K'$ valleys. This is the only case where a contribution with a minus sign is observed and interestingly enough is a contribution where the process involved is of the same order of $c_{1,0}=2$. This allows the edge states of the two $K$ and $K'$ valleys to mix with each other and makes a total of $2(c_{0,0}+c_{1,0})=3$, which is compatible with what is observed in Figs.~\ref{fig:loge} (a) and (c).

Figure~\ref{fig:loge} depicts the averaged local density of states near the edge of a semi-infinite plane for the radiated honeycomb lattice, using the recursive Green's-function method described in Ref.~\cite{Usaj2014}. Here we can also observe the higher-order gaps. Since the width of the gap is of order $\eta^{|n-m|}$ we use a logarithmic scale expanded around $\ve_0=0$ in Fig.~\ref{fig:loge} (c) and around $\ve_{1/2}=\hbar\Omega/2$ in Fig.~\ref{fig:loge} (d). This allows us to zoom in the spectrum up to a cut-off quasienergy denoted by $\tilde{\ve}$. This threshold is imposed arbitrarily, but constrained by the number of considered replicas and numerical precision. Note also that the weights of different replicas decay exponentially as $\eta^m$ for the $m\text{-th}$ replica; this is evident from the logarithmic scale in the color bar of Fig.~\ref{fig:loge}.

The procedure presented in this section accounts for the firsts orders of the generation of gaps and edge states and also has the advantage of retaining the largest gaps and the primary contributions to the averaged density of states. This procedure is correct if the quasienergies of the replicas involved lie within the van Hove singularities of each replica; otherwise, deviations due to the inaccuracy of the low-energy Dirac Hamiltonian appear and a full tight-binding model is required.

While there is a plethora of edge states appearing inside the gaps, some states might not be measurable simultaneously. In a transport experiment with non-irradiated leads only those which contribute significantly to the time averaged density of states will give a transport channel at the edge of the sample. In the approach of small $\eta$ the main contribution to the time-averaged density of states will be given only by the first-order gap and its associated edge state at $\ve\sim\hbar\Omega/2$, and to the second-order gap for $\ve\sim0$. For more details on the conductance for a transport calculation we refer the reader to~\cite{Foa2014}.
%
%
\subsection{\label{sec:hig}High intensity and high-frequency driving}
%
\begin{figure}[!tbp]
  \includegraphics[width=\columnwidth]{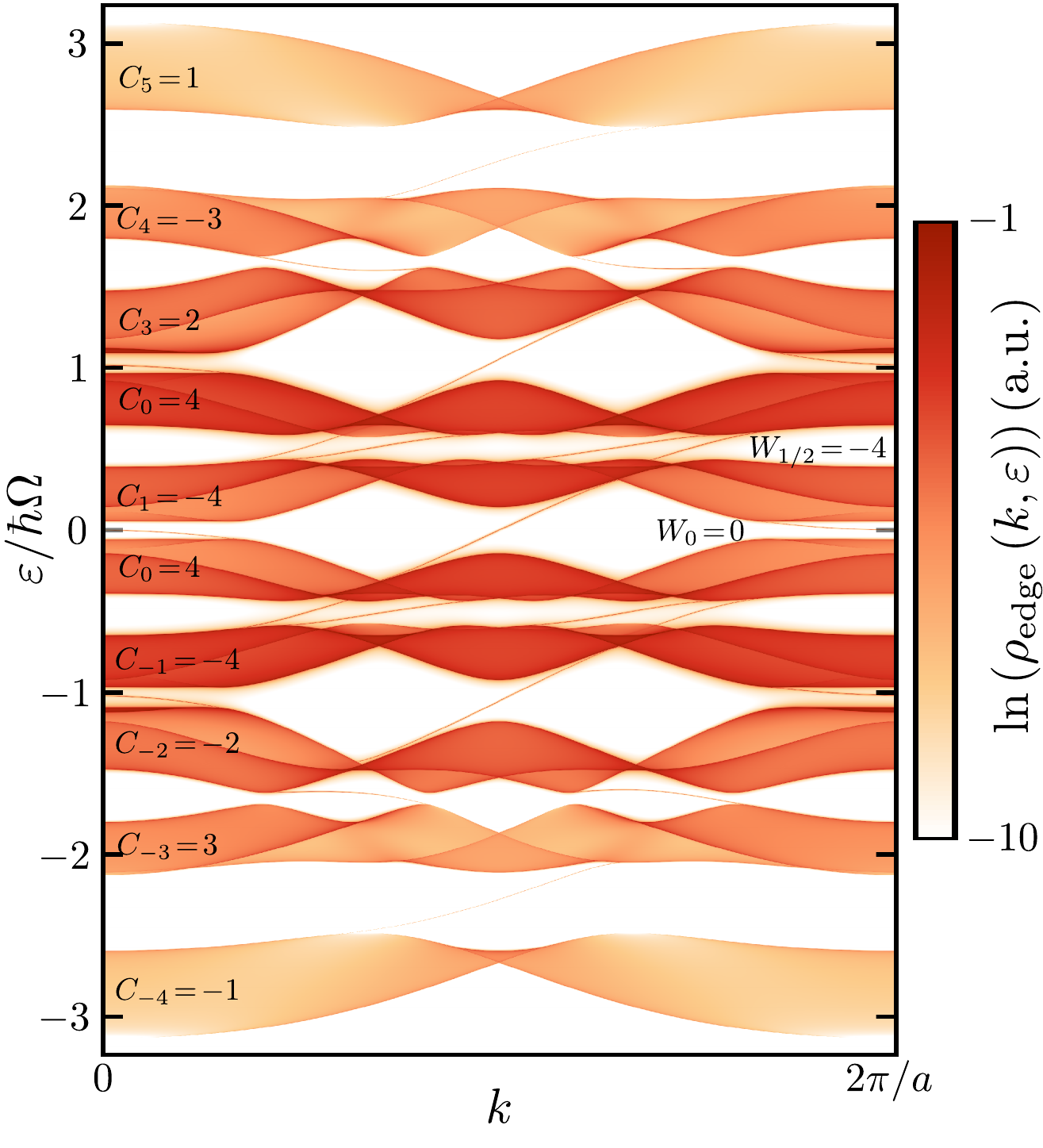}
  \caption{\label{fig:edge} Density of states near the edge of a semi-infinite honeycomb lattice. The lattice is driven under an electromagnetic field parametrized with frequency $\hbar\Omega=1.8\gamma$ and intensity $z=1.2$.}
\end{figure}
%
Now, let us briefly comment on the regime of high frequencies. Because of the reduced number of inelastic processes imposed by the higher energy cost, this regime is naturally less complex than the one addressed in the previous section. Notwithstanding, other difficulties must be taken care off. Indeed, for frequencies comparable to the band width, the low-energy approximation does not hold and the full tight-binding Hamiltonian is better suited in this case. For low intensities the system can still be solved perturbatively in the Floquet space or exactly for the truncated Floquet Hamiltonian, taking care of including at least all the Floquet replicas that fit in the replicas bandwidth, namely, $\Delta/\hw$, where $\Delta$ is the bandwidth [in our case $\Delta$ shrinks as $6\gamma J_0(z)$, where $z$ is the driving intensity, see Sec~\ref{sub:floquet}].

As the driving intensity is increased, higher-order inelastic processes are reinforced. Consequently the solutions for the infinite Floquet Hamiltonian are spread among more Floquet replicas. To obtain a numerical solution we truncate the Floquet Hamiltonian between the $-M$ and $M$ replicas. We must include as many replicas as needed for the winding number to converge. For example in Fig.~\ref{fig:edge}, even though $\Delta/\hw<3$, we need five Floquet replicas to obtain the correct result.

The construction of the winding number is also depicted in Fig.~\ref{fig:edge}, where each Floquet band has its associated Chern number at the left side, and the two relevant gaps at $\ve=0$ and $\hw/2$ have their associated winding numbers. The enhancement of the inelastic processes may lead to unexpected topological phase transitions as discussed in Sec.~\ref{sec:map}.
%
%
\section{\label{sec:map}Topological phase transitions}
%
\begin{figure*}[!tbp]
  \includegraphics[width=\textwidth]{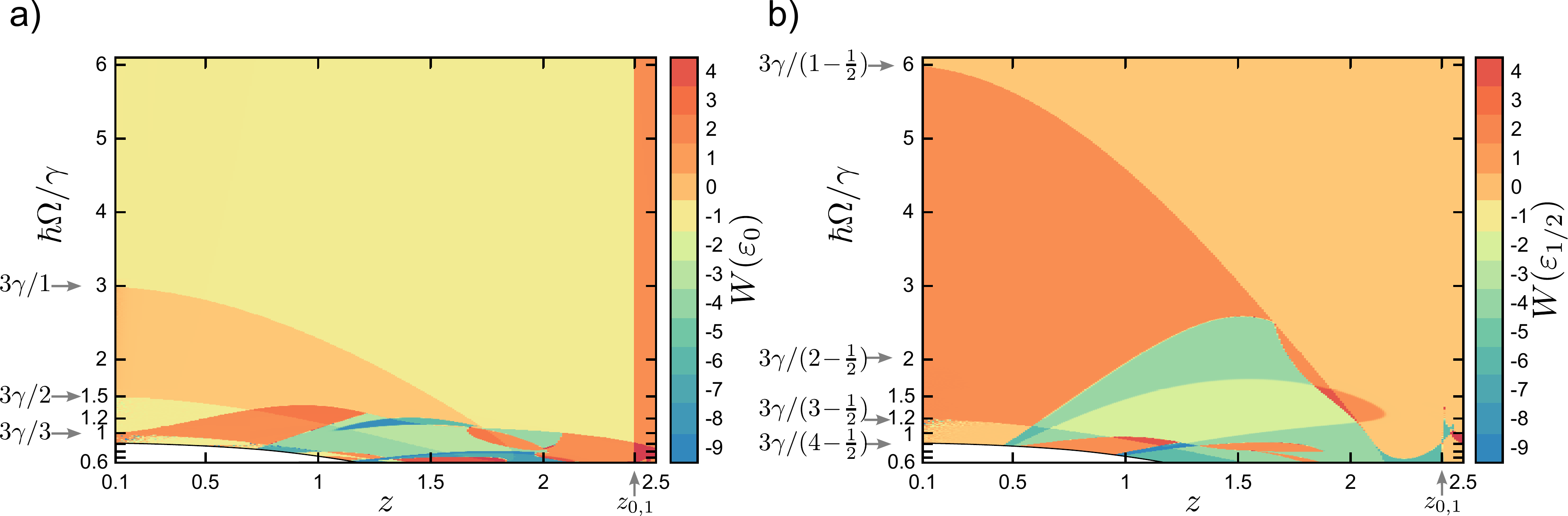}
  \caption{\label{fig:map} Map of the winding number $W(\ve)$, calculated with the full Floquet-Bloch bulk Hamiltonian, for $\varepsilon_0=0$ in panel (a) and $\varepsilon_{1/2}=\hbar\Omega/2$ in panel (b). A maximum number of $11$ replicas has been used throughout ($M=5$).
  }
\end{figure*}
%
In the previous section we showed that for low frequencies there is a growing number of edge states as a larger number of replicas are included in the calculation. There is, however, a natural limitation to this procedure, when the $\bm{k\,\cdot\,p}$ approach no longer describes correctly the topology of the Floquet bands involved. In the case of the honeycomb lattice the van Hove singularity sets this energy threshold.
The van Hove singularities lie at energies of $n\hbar\Omega\pm\gamma$ for the $n\text{-th}$ Floquet replica so it will be well described at the $\ve_0$ crossing only for frequencies such that $\hbar\Omega<\gamma/n$, and at the $\ve_{1/2}$ crossing for $\hbar\Omega<\gamma/(n-\frac{1}{2})$, assuming $n\ge 1$.
To illustrate this let us choose a frequency of $\hbar\Omega=0.05\gamma$. In this case the replicas from $m=-20$ to $20$ are well described at $\ve_0$, and the replicas $m=-18$ to $19$ are well described at $\ve_{1/2}$. So for low frequencies and moderate amplitudes the low-energy approach ensures to take into account all relevant Floquet replicas necessary for the calculation to converge, and to accurately address the number and chirality of the edge states relevant for transport.

As we increase the frequency, this number rapidly drops and one must use the full \emph{tight-binding} Hamiltonian to describe the bands in a wider energy range. To find out how many replicas are needed for the calculation of the Chern number to converge one must look at the replicas that would reach the $\ve_0$ and $\ve_{1/2}$ points for vanishing intensity ---for higher intensities more replicas are needed as explained below.

The bandwidth of the $n\text{-th}$ replica lies between $n\hbar\Omega\pm 3\gamma$, so it will be an overlap of different Floquet bands at frequencies $\hbar\Omega<3\gamma/n$ for the $\varepsilon_0$ crossing , and at $\hbar\Omega<3\gamma/(n-\frac{1}{2})$ for the $\ve_{1/2}$ crossing (assuming $n\ge 1$ and vanishing intensity). This behavior is shown in Fig.~\ref{fig:map}, where for low intensities a topological phase transition occurs every time a new pair of replicas enters in the description of the system. 
For low intensities, the bandwidth of the replica shrinks proportionally to $\gamma J_0(z)\approx\gamma(1-z^2)$, which can be seen as down going parabolas at $\hbar\Omega=3\gamma/n$ in Fig.~\ref{fig:map} (a) and at $\hbar\Omega<3\gamma/(n-\frac{1}{2})$ in Fig.~\ref{fig:map} (b).

The above deduction is based on the fact that for low intensities, the hoppings between one site in the $n\text{-th}$ replica and one site in the $m\text{-th}$ replica are proportional to $\gamma J_{n-m}(z)\sim\gamma z^{n-m}$. This means that for low intensities the dominant coupling is the zeroth-order one, i.e., the one within the same photon subspace. As one increases the intensity this assumption no longer holds and the coupling between neighboring replicas can achieve larger values, forcing the eigenfunctions that solve the Floquet Hamiltonian to be spread among many Floquet subspaces (replicas). 
For higher intensities the effects of introducing a new replica in the calculation extends beyond the replica's bandwidth and to correctly address the topology of the system one must include a larger number of replicas in the calculation. This is the explanation for the lines with a positive slope in Fig.~\ref{fig:map} that mark a topological phase transition.

Another interesting behavior is the transition at $z=2.4048$ for all frequencies, marked by a vertical line in Fig.~\ref{fig:map} (a) [and less resolved in (b)]. At this point the hopping between sites that belong to the same replica vanishes; this is the first zero $z_{0,1}$ of the Bessel function $J_0(x)$. Some numerical noise can be seen in panels (a) and (b) for low intensity because of the vanishing width of the highest-order gap considered, and also there is noise at some lines depicting a topological phase transition since the gap closes at every phase transition. The calculation time rapidly grows as more replicas are considered, which is the reason for the blank slices in the bottom left of (a) and (b). For larger values of $z$ a quasi periodic pattern is observed due to the Bessel's functions quasi-periodicity. This regime is not shown here because the intensities involved are extremely high for a possible experimental realization and for the assumptions made when modeling the electromagnetic field, and the system can become unstable against slight changes from circularly to elliptically polarized light, as studied in~\cite{Gomez-Leon2014} for high frequencies. Instead of the winding number, a map of the Chern number is presented in~\cite{Kundu2014}. Besides that, some phase transition could remain hidden for the time averaged transport in a multiterminal scattering configuration \cite{Foa2014}, since the corresponding edge states could bear no weight in the time-averaged density of states.
%
%
\section{Conclusions}
%
Characterizing the topological properties of driven systems in general, and honeycomb lattices in particular, is crucial for many studies pursuing novel Floquet topological phases~\cite{Lindner2011,Oka2009,Gomez-Leon2013,Goldman2014}. In this paper we address the calculation of the topological invariants in a wide range of parameters, from high to low frequencies. The Floquet quasienergy structure becomes progressively more complex when the frequency becomes much smaller than the bandwidth. In particular, within a small photon-energy range we find a nested structure of gaps of different widths, which are proportional to a power of the electron-photon coupling, the exponent being related to the order of the inelastic processes.

Interestingly, Floquet edge states develop within each gap even in the presence of a continuum of other Floquet bands provided that the edge states and the continuum have very different spectral weights among the replica's subspace. This allows one to devise a scheme for the determination of the number and chirality of the edge states where this information is progressively obtained as higher-order inelastic processes are included. This procedure is limited by the ratio between the system's bandwidth and the driving frequency.

The first stage of the scheme presented here is the calculation of an effective Hamiltonian, which is done analytically. This effective Hamiltonian is aimed at describing the Floquet quasienergy structure, rather than the time evolution, and allows one to compute the topological invariants in a broad set of driving frequencies and intensities. For low frequencies we have derived the contributions to the Chern numbers, constructively matching the numerical results obtained using recursive Green's functions for the Floquet-Bloch \emph{tight-binding} Hamiltonian.

For higher frequencies and a vast set of intensities the numerical evaluation of the winding number is summarized in a map of topological phase transitions. The main features are the lines that mark a topological phase transition where different numbers of Floquet replicas become degenerate. This allows one to tune the radiation parameters in order to obtain a specific number of edge states.
%
%
\section{Acknowledgements}
%
We thank Hern\'an L. Calvo and Carlos A. Balseiro for fruitful discussions. We acknowledge financial support from PICTs Grants No. 2008-2236, No. 2011-1552, and No. 2013-1045; Bicentenario Grant No. 2010-1060 from ANPCyT; PIP Grants No. 11220080101821 and No. 11220110100832 from CONICET and Grant No. 06/C415 from SeCyT-UNC. G.U. thanks the Simons Foundation, and L.E.F.F.T. the Alexander von Humboldt Foundation. G.U. and L.E.F.F.T. acknowledge support from the International Centre for Theoretical Physics (Trieste) associateship program.
%

\appendix*
%
\section{Effective Hamiltonian}
%
In this section we deduce the effective Hamiltonian that describes the crossing of different Floquet replicas, say the replicas labeled by $m$ and $n$. The crossings at $\ve_0$ occur for $m=-n$ and the crossings at $\ve_{1/2}$ occur for $m=-n+1$, where $n\ge1$. 
The most simple way to evaluate the effective coupling of the replicas is to make a change of basis that diagonalizes each subspace of the Floquet Hamiltonian $\mathcal{H}^\infty_{F}(\bm{k})$ in Eq.~\eqref{eq:Floquet_Ham_k} to get
%
\begin{equation}
  \begin{array}{l}
    \mathcal{\tilde{H}}^\infty_F(k,\theta_{\bm{k}}) = 
    \left( 
    \begin{array}{ccccc}
      \ddots & \vdots & \vdots & \vdots & \udots\\
      \cdots &  \tilde{H}_0^{(1)}(k) & V(\theta_{\bm{k}}) & 0 & \cdots\\
      \cdots & V(\theta_{\bm{k}})^\dagger & \tilde{H}_0^{(0)}(k) & V(\theta_{\bm{k}}) & \cdots\\
      \cdots & 0 & V(\theta_{\bm{k}})^\dagger & \tilde{H}_0^{(-1)}(k) & \cdots\\
      \udots & \vdots & \vdots & \vdots & \ddots
    \end{array} 
    \right)\,,
  \end{array}
  \label{matrix_k2}
\end{equation}
%
where
%
\begin{equation}
  \begin{array}{rcl}
    \tilde{H}_0^{(n) }(k)& = &\vf \left( 
    \begin{array}{cc}
      k & 0  \\
      0 & -k \\
    \end{array}
    \right)+n \vf 2 k_0\, I \\
    V(\theta_{\bm{k}}) &=& \eta k_0 \mathrm{e}^{-\ci \theta_k} \left(
    \begin{array}{cc}
       1 &  \mathrm{e}^{\ci \lambda}  \\
      -\mathrm{e}^{-\ci \lambda} & -1 \\
    \end{array}
    \right)\,,
  \end{array}
  \label{matrix_k3}
\end{equation}
%
where $\mathrm{e}^{\ci \lambda}$ is the trivial phase between the two basis vectors (in the following valued in $\lambda=0$), $\theta_{\bm{k}}$ is the angle between $\bm{k}$ and the $x$-axis, and $2 \vf k_0$ has replaced $\hbar\Omega$ to make evident the crossing point in the $\bm{k}$ space. As stated in Sec.~\ref{sec:low_energy} the values of the modulus of $\bm{k}$ where the replicas cross will be $k_{p,\beta}=2k_0(p+\beta)$, where $\beta=0$ or $1/2$, according to whether we are looking at the crossings at $\ve_0$ or $\ve_{1/2}$, respectively. 
The next step is to apply the decimation procedure, thus eliminating the replicas in between, to renormalize the effective hoppings that couple the desired replicas (the renormalization of the diagonal terms is irrelevant, for the purposes of calculating the Chern numbers, and will be neglected for simplicity). In the case that the replicas $m=0$ and $n=1$ it is straightforward to see from Eq.~\eqref{matrix_k2} that the effective Hamiltonian will be
%
\be
  \mathcal{H}_{F}^{\mathrm{eff}}(\bm{k},0,1/2)=\vf \left(
  \begin{array}{cc}
    -k+2k_0                             & -\eta k_0 \mathrm{e}^{-\ci \theta_k}\\
    -\eta k_0 \mathrm{e}^{\ci \theta_k}& k    
  \end{array}
  \right)
  \,,
\ee
%
since there are no replicas in between to decimate. At the same energy $\ve_{1/2}$ the next crossing will occur for the replicas $m=-1$ and $n=2$, and we will need two steps of decimation for the replicas zero and one. 

Then the decimation of two replicas will have the effect of accumulating two orders more in the coupling strength and in the phase factor, resulting in a coupling proportional to $\eta^3 \mathrm{e}^{-3\ci \theta_k}$. The calculation can be performed to easily obtain the effective Hamiltonians. Expressed in terms of $\bm{h}_{p,\beta}(\bm{k})$ the calculation yields,
%
\begin{eqnarray}
\nonumber
    \bm{h}_{0,1/2}(\bm{k})&=&\strut-\eta\,\hat{\bm{\xi}}_1\,k_{0,1/2}+ \left(-k+k_{0,1/2}\right)\,\hat{\bm{z}}\\
\nonumber
    \bm{h}_{1,1/2}(\bm{k})&\!\!=&-\eta^3\,\hat{\bm{\xi}}_3\,k_{1,1/2}\,\frac{3}{16} + \left(- k+k_{1,1/2}\right)\,\hat{\bm{z}}\\
\nonumber
    \bm{h}_{2,1/2}(\bm{k})&=&-\eta^5\,\hat{\bm{\xi}}_5\,k_{2,1/2}\,\frac{80}{3969} +\left(-k+k_{2,1/2}\right)\,\hat{\bm{z}}\\
    &&\dots\,\,\,,
 \end{eqnarray}
%
where the unit vector $\hat{\bm{\xi}}_n=\cos{(n\theta_{\bm{k}})}\,\hat{\bm{x}}+\sin{(n\theta_{\bm{k}})}\,\hat{\bm{y}}$ winds $n$ times in the $xy$ plane around the $z$ axis as we move $\bm{k}$, and $\mathcal{H}^{\mathrm{eff}}_F(\bm{k},p,\beta)=\vf \bm{h}_{p,\beta}(\bm{k})\cdot\bm{\sigma}+\ve_\beta \bm{I}$. Using this expressions we can evaluate Eq.~\eqref{c_rotado} to calculate the contribution of these crossings to the winding number, i.e., $c_{0,1/2}=1$, $c_{1,1/2}=3$, $c_{2,1/2}=5$, etc.

The same procedure can be applied to the crossings at $\ve_0$, starting from the crossing of the replicas $m=-1$ and $n=1$ at $k_{1,0}$, where only one decimation step is needed, giving a effective coupling proportional to $n^2\mathrm{e}^{-2\ci \theta_k}$. The next crossing of the replicas $m=-2$ and $n=2$ will accumulate two orders more in these factors, and so on. The explicit calculation gives
%
\begin{eqnarray}
\nonumber
    \bm{h}_{1,0}(\bm{k})&=&\eta^2\,\hat{\bm{\xi}}_2\, k_{1,0}\,\frac{ 1}{2}+\left(-k+k_{1,0}\right)\,\hat{\bm{z}}\\
\nonumber 
    \bm{h}_{2,0}(\bm{k})&=&\eta^4\,\hat{\bm{\xi}}_4\,k_{2,0}\,\frac{ 1}{18} +\left(-k+k_{2,0}\right)\,\hat{\bm{z}}\\
\nonumber
    \bm{h}_{3,0}(\bm{k})&=&\eta^6 \,\hat{\bm{\xi}}_6\,k_{3,0}\,\frac{9}{3200}  + \left(-k+k_{3,0}\right)\hat{\bm{z}}\\
    &&\dots\,\,\,,
\end{eqnarray}
%
where $\bm{h}_{p,\beta}(\bm{k})$ is defined as before. It is straight forward to see that $c_{1,0}=2$, $c_{2,0}=4$, $c_{3,0}=6$, etc.

The only exception is the calculation of $\bm{h}_{0,0}$, which has been already addressed by Oka and Aoki~\cite{Oka2009}. This time, the effective Hamiltonian is the renormalized Hamiltonian of the $m=n=0$ replica which has a crossing of its own bands at the Dirac point $k_{0,0}=0$. The degeneracy is lifted due to the coupling with the replicas $\pm1$, and the effective Hamiltonian is described by Eq.~\eqref{eq:h_00}, and can be equally expressed as,
\be
\bm{h}_{0,0}(\bm{k}) = -\eta^2\, 2 k_0\hat{\bm{x}} +k\hat{\bm{z}}\,.
\ee
In this case, it is important to rotate back to a $k$-independent basis as explained in the text. This is done with the rotation matrix
\be
\bm{R}(\theta)=
\left(
  \begin{array}{ccc}
 0 & \sin\theta &\cos\theta\\
0 & -\cos\theta & \sin\theta\\
1 & 0 & 0
  \end{array}
  \right)\,,
\ee
that satisfies the following useful identity
\be
\bm{R}^{-1}\partial_\theta\bm{R}\, \hat{\bm{h}}_{p,\beta}=\hat{\bm{x}}\times\hat{\bm{h}}_{p,\beta}\,.
\ee

\bibliographystyle{apsrev4-1_title}
\bibliography{ac-graphene}

\begin{thebibliography}{56}%
\makeatletter
\providecommand \@ifxundefined [1]{%
 \@ifx{#1\undefined}
}%
\providecommand \@ifnum [1]{%
 \ifnum #1\expandafter \@firstoftwo
 \else \expandafter \@secondoftwo
 \fi
}%
\providecommand \@ifx [1]{%
 \ifx #1\expandafter \@firstoftwo
 \else \expandafter \@secondoftwo
 \fi
}%
\providecommand \natexlab [1]{#1}%
\providecommand \enquote  [1]{``#1''}%
\providecommand \bibnamefont  [1]{#1}%
\providecommand \bibfnamefont [1]{#1}%
\providecommand \citenamefont [1]{#1}%
\providecommand \href@noop [0]{\@secondoftwo}%
\providecommand \href [0]{\begingroup \@sanitize@url \@href}%
\providecommand \@href[1]{\@@startlink{#1}\@@href}%
\providecommand \@@href[1]{\endgroup#1\@@endlink}%
\providecommand \@sanitize@url [0]{\catcode `\\12\catcode `\$12\catcode
  `\&12\catcode `\#12\catcode `\^12\catcode `\_12\catcode `\%12\relax}%
\providecommand \@@startlink[1]{}%
\providecommand \@@endlink[0]{}%
\providecommand \url  [0]{\begingroup\@sanitize@url \@url }%
\providecommand \@url [1]{\endgroup\@href {#1}{\urlprefix }}%
\providecommand \urlprefix  [0]{URL }%
\providecommand \Eprint [0]{\href }%
\providecommand \doibase [0]{http://dx.doi.org/}%
\providecommand \selectlanguage [0]{\@gobble}%
\providecommand \bibinfo  [0]{\@secondoftwo}%
\providecommand \bibfield  [0]{\@secondoftwo}%
\providecommand \translation [1]{[#1]}%
\providecommand \BibitemOpen [0]{}%
\providecommand \bibitemStop [0]{}%
\providecommand \bibitemNoStop [0]{.\EOS\space}%
\providecommand \EOS [0]{\spacefactor3000\relax}%
\providecommand \BibitemShut  [1]{\csname bibitem#1\endcsname}%
\let\auto@bib@innerbib\@empty
\bibitem [{\citenamefont {Hasan}\ and\ \citenamefont {Kane}(2010)}]{Hasan2010}%
  \BibitemOpen
  \bibfield  {author} {\bibinfo {author} {\bibfnamefont {M.~Z.}\ \bibnamefont
  {Hasan}}\ and\ \bibinfo {author} {\bibfnamefont {C.~L.}\ \bibnamefont
  {Kane}},\ }\bibfield  {title} {\enquote {\bibinfo {title}
  {\textit{Colloquium} : Topological insulators},}\ }\href {\doibase
  10.1103/RevModPhys.82.3045} {\bibfield  {journal} {\bibinfo  {journal} {Rev.
  Mod. Phys.}\ }\textbf {\bibinfo {volume} {82}},\ \bibinfo {pages} {3045}
  (\bibinfo {year} {2010})}\BibitemShut {NoStop}%
\bibitem [{\citenamefont {Bernevig}\ and\ \citenamefont
  {Hughes}(2013)}]{Bernevig2013}%
  \BibitemOpen
  \bibfield  {author} {\bibinfo {author} {\bibfnamefont {B.~A.}\ \bibnamefont
  {Bernevig}}\ and\ \bibinfo {author} {\bibfnamefont {T.~L.}\ \bibnamefont
  {Hughes}},\ }\href@noop {} {\emph {\bibinfo {title} {Topological Insulators
  and Topological Superconductors}}}\ (\bibinfo  {publisher} {Princeton
  University Press},\ \bibinfo {year} {2013})\BibitemShut {NoStop}%
\bibitem [{\citenamefont {Oka}\ and\ \citenamefont {Aoki}(2009)}]{Oka2009}%
  \BibitemOpen
  \bibfield  {author} {\bibinfo {author} {\bibfnamefont {T.}~\bibnamefont
  {Oka}}\ and\ \bibinfo {author} {\bibfnamefont {H.}~\bibnamefont {Aoki}},\
  }\bibfield  {title} {\enquote {\bibinfo {title} {Photovoltaic hall effect in
  graphene},}\ }\href {http://link.aps.org/doi/10.1103/PhysRevB.79.081406}
  {\bibfield  {journal} {\bibinfo  {journal} {Phys. Rev. B}\ }\textbf {\bibinfo
  {volume} {79}},\ \bibinfo {pages} {081406} (\bibinfo {year}
  {2009})}\BibitemShut {NoStop}%
\bibitem [{\citenamefont {Kitagawa}\ \emph {et~al.}(2010)\citenamefont
  {Kitagawa}, \citenamefont {Berg}, \citenamefont {Rudner},\ and\ \citenamefont
  {Demler}}]{Kitagawa2010}%
  \BibitemOpen
  \bibfield  {author} {\bibinfo {author} {\bibfnamefont {T.}~\bibnamefont
  {Kitagawa}}, \bibinfo {author} {\bibfnamefont {E.}~\bibnamefont {Berg}},
  \bibinfo {author} {\bibfnamefont {M.}~\bibnamefont {Rudner}}, \ and\ \bibinfo
  {author} {\bibfnamefont {E.}~\bibnamefont {Demler}},\ }\bibfield  {title}
  {\enquote {\bibinfo {title} {Topological characterization of periodically
  driven quantum systems},}\ }\href
  {http://link.aps.org/doi/10.1103/PhysRevB.82.235114} {\bibfield  {journal}
  {\bibinfo  {journal} {Phys. Rev. B}\ }\textbf {\bibinfo {volume} {82}},\
  \bibinfo {pages} {235114} (\bibinfo {year} {2010})}\BibitemShut {NoStop}%
\bibitem [{\citenamefont {Lindner}\ \emph {et~al.}(2011)\citenamefont
  {Lindner}, \citenamefont {Refael},\ and\ \citenamefont
  {Galitski}}]{Lindner2011}%
  \BibitemOpen
  \bibfield  {author} {\bibinfo {author} {\bibfnamefont {N.~H.}\ \bibnamefont
  {Lindner}}, \bibinfo {author} {\bibfnamefont {G.}~\bibnamefont {Refael}}, \
  and\ \bibinfo {author} {\bibfnamefont {V.}~\bibnamefont {Galitski}},\
  }\bibfield  {title} {\enquote {\bibinfo {title} {Floquet topological
  insulator in semiconductor quantum wells},}\ }\href
  {http://dx.doi.org/10.1038/nphys1926} {\bibfield  {journal} {\bibinfo
  {journal} {Nat Phys}\ }\textbf {\bibinfo {volume} {7}},\ \bibinfo {pages}
  {490} (\bibinfo {year} {2011})}\BibitemShut {NoStop}%
\bibitem [{\citenamefont {Calvo}\ \emph {et~al.}(2011)\citenamefont {Calvo},
  \citenamefont {Pastawski}, \citenamefont {Roche},\ and\ \citenamefont
  {Foa~Torres}}]{Calvo2011}%
  \BibitemOpen
  \bibfield  {author} {\bibinfo {author} {\bibfnamefont {H.~L.}\ \bibnamefont
  {Calvo}}, \bibinfo {author} {\bibfnamefont {H.~M.}\ \bibnamefont
  {Pastawski}}, \bibinfo {author} {\bibfnamefont {S.}~\bibnamefont {Roche}}, \
  and\ \bibinfo {author} {\bibfnamefont {L.~E.~F.}\ \bibnamefont
  {Foa~Torres}},\ }\bibfield  {title} {\enquote {\bibinfo {title} {Tuning
  laser-induced band gaps in graphene},}\ }\href
  {http://dx.doi.org/10.1063/1.3597412} {\bibfield  {journal} {\bibinfo
  {journal} {Appl. Phys. Lett.}\ }\textbf {\bibinfo {volume} {98}},\ \bibinfo
  {pages} {232103} (\bibinfo {year} {2011})}\BibitemShut {NoStop}%
\bibitem [{\citenamefont {Zhou}\ and\ \citenamefont {Wu}(2011)}]{Zhou2011}%
  \BibitemOpen
  \bibfield  {author} {\bibinfo {author} {\bibfnamefont {Y.}~\bibnamefont
  {Zhou}}\ and\ \bibinfo {author} {\bibfnamefont {M.~W.}\ \bibnamefont {Wu}},\
  }\bibfield  {title} {\enquote {\bibinfo {title} {Optical response of graphene
  under intense terahertz fields},}\ }\href
  {http://link.aps.org/doi/10.1103/PhysRevB.83.245436} {\bibfield  {journal}
  {\bibinfo  {journal} {Phys. Rev. B}\ }\textbf {\bibinfo {volume} {83}},\
  \bibinfo {pages} {245436} (\bibinfo {year} {2011})}\BibitemShut {NoStop}%
\bibitem [{\citenamefont {Kitagawa}\ \emph {et~al.}(2011)\citenamefont
  {Kitagawa}, \citenamefont {Oka}, \citenamefont {Brataas}, \citenamefont
  {Fu},\ and\ \citenamefont {Demler}}]{Kitagawa2011}%
  \BibitemOpen
  \bibfield  {author} {\bibinfo {author} {\bibfnamefont {T.}~\bibnamefont
  {Kitagawa}}, \bibinfo {author} {\bibfnamefont {T.}~\bibnamefont {Oka}},
  \bibinfo {author} {\bibfnamefont {A.}~\bibnamefont {Brataas}}, \bibinfo
  {author} {\bibfnamefont {L.}~\bibnamefont {Fu}}, \ and\ \bibinfo {author}
  {\bibfnamefont {E.}~\bibnamefont {Demler}},\ }\bibfield  {title} {\enquote
  {\bibinfo {title} {Transport properties of nonequilibrium systems under the
  application of light: Photoinduced quantum hall insulators without landau
  levels},}\ }\href {http://link.aps.org/doi/10.1103/PhysRevB.84.235108}
  {\bibfield  {journal} {\bibinfo  {journal} {Phys. Rev. B}\ }\textbf {\bibinfo
  {volume} {84}},\ \bibinfo {pages} {235108} (\bibinfo {year}
  {2011})}\BibitemShut {NoStop}%
\bibitem [{\citenamefont {Iurov}\ \emph {et~al.}(2012)\citenamefont {Iurov},
  \citenamefont {Gumbs}, \citenamefont {Roslyak},\ and\ \citenamefont
  {Huang}}]{Iurov2012}%
  \BibitemOpen
  \bibfield  {author} {\bibinfo {author} {\bibfnamefont {A.}~\bibnamefont
  {Iurov}}, \bibinfo {author} {\bibfnamefont {G.}~\bibnamefont {Gumbs}},
  \bibinfo {author} {\bibfnamefont {O.}~\bibnamefont {Roslyak}}, \ and\
  \bibinfo {author} {\bibfnamefont {D.}~\bibnamefont {Huang}},\ }\bibfield
  {title} {\enquote {\bibinfo {title} {Anomalous photon-assisted tunneling in
  graphene},}\ }\href {http://stacks.iop.org/0953-8984/24/i=1/a=015303}
  {\bibfield  {journal} {\bibinfo  {journal} {Journal of Physics: Condensed
  Matter}\ }\textbf {\bibinfo {volume} {24}},\ \bibinfo {pages} {015303}
  (\bibinfo {year} {2012})}\BibitemShut {NoStop}%
\bibitem [{\citenamefont {Su\'arez~Morell}\ and\ \citenamefont
  {Foa~Torres}(2012)}]{SuarezMorell2012}%
  \BibitemOpen
  \bibfield  {author} {\bibinfo {author} {\bibfnamefont {E.}~\bibnamefont
  {Su\'arez~Morell}}\ and\ \bibinfo {author} {\bibfnamefont {L.~E.~F.}\
  \bibnamefont {Foa~Torres}},\ }\bibfield  {title} {\enquote {\bibinfo {title}
  {Radiation effects on the electronic properties of bilayer graphene},}\
  }\href {\doibase 10.1103/PhysRevB.86.125449} {\bibfield  {journal} {\bibinfo
  {journal} {Phys. Rev. B}\ }\textbf {\bibinfo {volume} {86}},\ \bibinfo
  {pages} {125449} (\bibinfo {year} {2012})}\BibitemShut {NoStop}%
\bibitem [{\citenamefont {Perez-Piskunow}\ \emph {et~al.}(2014)\citenamefont
  {Perez-Piskunow}, \citenamefont {Usaj}, \citenamefont {Balseiro},\ and\
  \citenamefont {Foa~Torres}}]{Perez-Piskunow2014}%
  \BibitemOpen
  \bibfield  {author} {\bibinfo {author} {\bibfnamefont {P.~M.}\ \bibnamefont
  {Perez-Piskunow}}, \bibinfo {author} {\bibfnamefont {G.}~\bibnamefont
  {Usaj}}, \bibinfo {author} {\bibfnamefont {C.~A.}\ \bibnamefont {Balseiro}},
  \ and\ \bibinfo {author} {\bibfnamefont {L.~E.~F.}\ \bibnamefont
  {Foa~Torres}},\ }\bibfield  {title} {\enquote {\bibinfo {title} {Floquet
  chiral edge states in graphene},}\ }\href {\doibase
  10.1103/PhysRevB.89.121401} {\bibfield  {journal} {\bibinfo  {journal} {Phys.
  Rev. B}\ }\textbf {\bibinfo {volume} {89}},\ \bibinfo {pages} {121401(R)}
  (\bibinfo {year} {2014})}\BibitemShut {NoStop}%
\bibitem [{\citenamefont {Usaj}\ \emph {et~al.}(2014)\citenamefont {Usaj},
  \citenamefont {Perez-Piskunow}, \citenamefont {Foa~Torres},\ and\
  \citenamefont {Balseiro}}]{Usaj2014}%
  \BibitemOpen
  \bibfield  {author} {\bibinfo {author} {\bibfnamefont {G.}~\bibnamefont
  {Usaj}}, \bibinfo {author} {\bibfnamefont {P.~M.}\ \bibnamefont
  {Perez-Piskunow}}, \bibinfo {author} {\bibfnamefont {L.~E.~F.}\ \bibnamefont
  {Foa~Torres}}, \ and\ \bibinfo {author} {\bibfnamefont {C.~A.}\ \bibnamefont
  {Balseiro}},\ }\bibfield  {title} {\enquote {\bibinfo {title} {Irradiated
  graphene as a tunable floquet topological insulator},}\ }\href {\doibase
  10.1103/PhysRevB.90.115423} {\bibfield  {journal} {\bibinfo  {journal} {Phys.
  Rev. B}\ }\textbf {\bibinfo {volume} {90}},\ \bibinfo {pages} {115423}
  (\bibinfo {year} {2014})}\BibitemShut {NoStop}%
\bibitem [{\citenamefont {Sie}\ \emph {et~al.}(2014)\citenamefont {Sie},
  \citenamefont {McIver}, \citenamefont {Lee}, \citenamefont {Fu},
  \citenamefont {Kong},\ and\ \citenamefont {Gedik}}]{Sie2015}%
  \BibitemOpen
  \bibfield  {author} {\bibinfo {author} {\bibfnamefont {E.~J.}\ \bibnamefont
  {Sie}}, \bibinfo {author} {\bibfnamefont {J.~W.}\ \bibnamefont {McIver}},
  \bibinfo {author} {\bibfnamefont {Y.-H.}\ \bibnamefont {Lee}}, \bibinfo
  {author} {\bibfnamefont {L.}~\bibnamefont {Fu}}, \bibinfo {author}
  {\bibfnamefont {J.}~\bibnamefont {Kong}}, \ and\ \bibinfo {author}
  {\bibfnamefont {N.}~\bibnamefont {Gedik}},\ }\bibfield  {title} {\enquote
  {\bibinfo {title} {Valley-selective optical stark effect in
  monolayer ws2},}\ }\href {http://dx.doi.org/10.1038/nmat4156} {\bibfield
  {journal} {\bibinfo  {journal} {Nature Materials}\ }\textbf {\bibinfo
  {volume} {14}},\ \bibinfo {pages} {290} (\bibinfo {year} {2014})}\BibitemShut
  {NoStop}%
\bibitem [{\citenamefont {L\'opez}\ \emph {et~al.}(2015)\citenamefont
  {L\'opez}, \citenamefont {Scholz}, \citenamefont {Santos},\ and\
  \citenamefont {Schliemann}}]{Lopez2015}%
  \BibitemOpen
  \bibfield  {author} {\bibinfo {author} {\bibfnamefont {A.}~\bibnamefont
  {L\'opez}}, \bibinfo {author} {\bibfnamefont {A.}~\bibnamefont {Scholz}},
  \bibinfo {author} {\bibfnamefont {B.}~\bibnamefont {Santos}}, \ and\ \bibinfo
  {author} {\bibfnamefont {J.}~\bibnamefont {Schliemann}},\ }\bibfield  {title}
  {\enquote {\bibinfo {title} {Photoinduced pseudospin effects in silicene
  beyond the off resonant condition},}\ }\href {\doibase
  http://dx.doi.org/10.1103/PhysRevB.91.125105} {\bibfield  {journal} {\bibinfo
   {journal} {Phys. Rev. B}\ }\textbf {\bibinfo {volume} {91}},\ \bibinfo
  {pages} {125105} (\bibinfo {year} {2015})}\BibitemShut {NoStop}%
\bibitem [{\citenamefont {D\'ora}\ \emph {et~al.}(2012)\citenamefont {D\'ora},
  \citenamefont {Cayssol}, \citenamefont {Simon},\ and\ \citenamefont
  {Moessner}}]{Dora2012}%
  \BibitemOpen
  \bibfield  {author} {\bibinfo {author} {\bibfnamefont {B.}~\bibnamefont
  {D\'ora}}, \bibinfo {author} {\bibfnamefont {J.}~\bibnamefont {Cayssol}},
  \bibinfo {author} {\bibfnamefont {F.}~\bibnamefont {Simon}}, \ and\ \bibinfo
  {author} {\bibfnamefont {R.}~\bibnamefont {Moessner}},\ }\bibfield  {title}
  {\enquote {\bibinfo {title} {Optically engineering the topological properties
  of a spin hall insulator},}\ }\href
  {http://link.aps.org/doi/10.1103/PhysRevLett.108.056602} {\bibfield
  {journal} {\bibinfo  {journal} {Phys. Rev. Lett.}\ }\textbf {\bibinfo
  {volume} {108}},\ \bibinfo {pages} {056602} (\bibinfo {year}
  {2012})}\BibitemShut {NoStop}%
\bibitem [{\citenamefont {Calvo}\ \emph {et~al.}()\citenamefont {Calvo},
  \citenamefont {Foa~Torres}, \citenamefont {Perez-Piskunow}, \citenamefont
  {Balseiro},\ and\ \citenamefont {Usaj}}]{Calvo2015}%
  \BibitemOpen
  \bibfield  {author} {\bibinfo {author} {\bibfnamefont {H.~L.}\ \bibnamefont
  {Calvo}}, \bibinfo {author} {\bibfnamefont {L.~E.~F.}\ \bibnamefont
  {Foa~Torres}}, \bibinfo {author} {\bibfnamefont {P.~M.}\ \bibnamefont
  {Perez-Piskunow}}, \bibinfo {author} {\bibfnamefont {C.~A.}\ \bibnamefont
  {Balseiro}}, \ and\ \bibinfo {author} {\bibfnamefont {G.}~\bibnamefont
  {Usaj}},\ }\bibfield  {title} {\enquote {\bibinfo {title} {Floquet interface
  states in illuminated three dimensional topological insulators},}\ }\href
  {http://arxiv.org/abs/1502.04098} {\bibinfo  {journal} {arXiv:1502.04098
  [cond-mat.mes-hall]}\ }\BibitemShut {NoStop}%
\bibitem [{\citenamefont {Rechtsman}\ \emph {et~al.}(2013)\citenamefont
  {Rechtsman}, \citenamefont {Zeuner}, \citenamefont {Plotnik}, \citenamefont
  {Lumer}, \citenamefont {Podolsky}, \citenamefont {Dreisow}, \citenamefont
  {Nolte}, \citenamefont {Segev},\ and\ \citenamefont
  {Szameit}}]{Rechtsman2013}%
  \BibitemOpen
\bibfield  {journal} {  }\bibfield  {author} {\bibinfo {author} {\bibfnamefont
  {M.~C.}\ \bibnamefont {Rechtsman}}, \bibinfo {author} {\bibfnamefont {J.~M.}\
  \bibnamefont {Zeuner}}, \bibinfo {author} {\bibfnamefont {Y.}~\bibnamefont
  {Plotnik}}, \bibinfo {author} {\bibfnamefont {Y.}~\bibnamefont {Lumer}},
  \bibinfo {author} {\bibfnamefont {D.}~\bibnamefont {Podolsky}}, \bibinfo
  {author} {\bibfnamefont {F.}~\bibnamefont {Dreisow}}, \bibinfo {author}
  {\bibfnamefont {S.}~\bibnamefont {Nolte}}, \bibinfo {author} {\bibfnamefont
  {M.}~\bibnamefont {Segev}}, \ and\ \bibinfo {author} {\bibfnamefont
  {A.}~\bibnamefont {Szameit}},\ }\bibfield  {title} {\enquote {\bibinfo
  {title} {Photonic floquet topological insulators},}\ }\href
  {http://dx.doi.org/10.1038/nature12066} {\bibfield  {journal} {\bibinfo
  {journal} {Nature}\ }\textbf {\bibinfo {volume} {496}},\ \bibinfo {pages}
  {196} (\bibinfo {year} {2013})}\BibitemShut {NoStop}%
\bibitem [{\citenamefont {Goldman}\ and\ \citenamefont
  {Dalibard}(2014)}]{Goldman2014}%
  \BibitemOpen
  \bibfield  {author} {\bibinfo {author} {\bibfnamefont {N.}~\bibnamefont
  {Goldman}}\ and\ \bibinfo {author} {\bibfnamefont {J.}~\bibnamefont
  {Dalibard}},\ }\bibfield  {title} {\enquote {\bibinfo {title} {Periodically
  driven quantum systems: Effective hamiltonians and engineered gauge
  fields},}\ }\href {\doibase 10.1103/PhysRevX.4.031027} {\bibfield  {journal}
  {\bibinfo  {journal} {Phys. Rev. X}\ }\textbf {\bibinfo {volume} {4}},\
  \bibinfo {pages} {031027} (\bibinfo {year} {2014})}\BibitemShut {NoStop}%
\bibitem [{\citenamefont {Choudhury}\ and\ \citenamefont
  {Mueller}(2014)}]{Choudhury2014}%
  \BibitemOpen
  \bibfield  {author} {\bibinfo {author} {\bibfnamefont {S.}~\bibnamefont
  {Choudhury}}\ and\ \bibinfo {author} {\bibfnamefont {E.~J.}\ \bibnamefont
  {Mueller}},\ }\bibfield  {title} {\enquote {\bibinfo {title} {Stability of a
  floquet bose-einstein condensate in a one-dimensional optical lattice},}\
  }\href {\doibase http://dx.doi.org/10.1103/PhysRevA.90.013621} {\bibfield
  {journal} {\bibinfo  {journal} {Phys. Rev. A}\ }\textbf {\bibinfo {volume}
  {90}},\ \bibinfo {pages} {013621} (\bibinfo {year} {2014})}\BibitemShut
  {NoStop}%
\bibitem [{\citenamefont {Bukov}\ \emph {et~al.}()\citenamefont {Bukov},
  \citenamefont {D'Alessio},\ and\ \citenamefont {Polkovnikov}}]{Bukov2014}%
  \BibitemOpen
  \bibfield  {author} {\bibinfo {author} {\bibfnamefont {M.}~\bibnamefont
  {Bukov}}, \bibinfo {author} {\bibfnamefont {L.}~\bibnamefont {D'Alessio}}, \
  and\ \bibinfo {author} {\bibfnamefont {A.}~\bibnamefont {Polkovnikov}},\
  }\bibfield  {title} {\enquote {\bibinfo {title} {Universal high-frequency
  behavior of periodically driven systems: from dynamical stabilization to
  floquet engineering},}\ }\href {http://arxiv.org/abs/1407.4803} {\bibinfo
  {journal} {arXiv:1407.4803 [cond-mat.quant-gas]}\ }\BibitemShut {NoStop}%
\bibitem [{\citenamefont {Bilitewski}\ and\ \citenamefont
  {Cooper}(2015)}]{Bilitewski2014}%
  \BibitemOpen
\bibfield  {journal} {  }\bibfield  {author} {\bibinfo {author} {\bibfnamefont
  {T.}~\bibnamefont {Bilitewski}}\ and\ \bibinfo {author} {\bibfnamefont
  {N.~R.}\ \bibnamefont {Cooper}},\ }\bibfield  {title} {\enquote {\bibinfo
  {title} {Scattering theory for floquet-bloch states},}\ }\href {\doibase
  10.1103/PhysRevA.91.033601} {\bibfield  {journal} {\bibinfo  {journal} {Phys.
  Rev. A}\ }\textbf {\bibinfo {volume} {91}},\ \bibinfo {pages} {033601}
  (\bibinfo {year} {2015})}\BibitemShut {NoStop}%
\bibitem [{\citenamefont {Dasgupta}\ \emph {et~al.}()\citenamefont {Dasgupta},
  \citenamefont {Bhattacharya},\ and\ \citenamefont {Dutta}}]{Dasgupta2015}%
  \BibitemOpen
  \bibfield  {author} {\bibinfo {author} {\bibfnamefont {S.}~\bibnamefont
  {Dasgupta}}, \bibinfo {author} {\bibfnamefont {U.}~\bibnamefont
  {Bhattacharya}}, \ and\ \bibinfo {author} {\bibfnamefont {A.}~\bibnamefont
  {Dutta}},\ }\bibfield  {title} {\enquote {\bibinfo {title} {Phase transition
  in the periodically pulsed dicke model},}\ }\href
  {http://arxiv.org/abs/1412.6460} {\bibinfo  {journal} {arXiv:1412.6460
  [cond-mat.stat-mech]}\ }\BibitemShut {NoStop}%
\bibitem [{\citenamefont {Tong}\ \emph {et~al.}(2013)\citenamefont {Tong},
  \citenamefont {An}, \citenamefont {Gong}, \citenamefont {Luo},\ and\
  \citenamefont {Oh}}]{Qing2013}%
  \BibitemOpen
\bibfield  {journal} {  }\bibfield  {author} {\bibinfo {author} {\bibfnamefont
  {Q.-J.}\ \bibnamefont {Tong}}, \bibinfo {author} {\bibfnamefont {J.-H.}\
  \bibnamefont {An}}, \bibinfo {author} {\bibfnamefont {J.}~\bibnamefont
  {Gong}}, \bibinfo {author} {\bibfnamefont {H.-G.}\ \bibnamefont {Luo}}, \
  and\ \bibinfo {author} {\bibfnamefont {C.~H.}\ \bibnamefont {Oh}},\
  }\bibfield  {title} {\enquote {\bibinfo {title} {Generating many majorana
  modes via periodic driving: A superconductor model},}\ }\href {\doibase
  10.1103/PhysRevB.87.201109} {\bibfield  {journal} {\bibinfo  {journal} {Phys.
  Rev. B}\ }\textbf {\bibinfo {volume} {87}},\ \bibinfo {pages} {201109}
  (\bibinfo {year} {2013})}\BibitemShut {NoStop}%
\bibitem [{\citenamefont {D'Alessio}\ and\ \citenamefont
  {Rigol}()}]{DAlessio2014}%
  \BibitemOpen
  \bibfield  {author} {\bibinfo {author} {\bibfnamefont {L.}~\bibnamefont
  {D'Alessio}}\ and\ \bibinfo {author} {\bibfnamefont {M.}~\bibnamefont
  {Rigol}},\ }\bibfield  {title} {\enquote {\bibinfo {title} {Dynamical
  preparation of floquet chern insulators: A no-go theorem and the
  experiments},}\ }\href {http://arxiv.org/abs/1409.6319} {\bibinfo  {journal}
  {arXiv:1409.6319 [cond-mat.quant-gas]}\ }\BibitemShut {NoStop}%
\bibitem [{\citenamefont {Goldman}\ \emph {et~al.}(2015)\citenamefont
  {Goldman}, \citenamefont {Dalibard}, \citenamefont {Aidelsburger},\ and\
  \citenamefont {Cooper}}]{Goldman2015}%
  \BibitemOpen
\bibfield  {journal} {  }\bibfield  {author} {\bibinfo {author} {\bibfnamefont
  {N.}~\bibnamefont {Goldman}}, \bibinfo {author} {\bibfnamefont
  {J.}~\bibnamefont {Dalibard}}, \bibinfo {author} {\bibfnamefont
  {M.}~\bibnamefont {Aidelsburger}}, \ and\ \bibinfo {author} {\bibfnamefont
  {N.~R.}\ \bibnamefont {Cooper}},\ }\bibfield  {title} {\enquote {\bibinfo
  {title} {Periodically driven quantum matter: The case of resonant
  modulations},}\ }\href {\doibase 10.1103/PhysRevA.91.033632} {\bibfield
  {journal} {\bibinfo  {journal} {Phys. Rev. A}\ }\textbf {\bibinfo {volume}
  {91}},\ \bibinfo {pages} {033632} (\bibinfo {year} {2015})}\BibitemShut
  {NoStop}%
\bibitem [{\citenamefont {Mori}(2015)}]{Mori2015}%
  \BibitemOpen
  \bibfield  {author} {\bibinfo {author} {\bibfnamefont {T.}~\bibnamefont
  {Mori}},\ }\bibfield  {title} {\enquote {\bibinfo {title} {Floquet resonant
  states and validity of the floquet-magnus expansion in the periodically
  driven friedrichs models},}\ }\href {\doibase 10.1103/PhysRevA.91.020101}
  {\bibfield  {journal} {\bibinfo  {journal} {Phys. Rev. A}\ }\textbf {\bibinfo
  {volume} {91}},\ \bibinfo {pages} {020101} (\bibinfo {year}
  {2015})}\BibitemShut {NoStop}%
\bibitem [{\citenamefont {Dahlhaus}\ \emph {et~al.}()\citenamefont {Dahlhaus},
  \citenamefont {Fregoso},\ and\ \citenamefont {Moore}}]{Dahlhaus2014}%
  \BibitemOpen
  \bibfield  {author} {\bibinfo {author} {\bibfnamefont {J.~P.}\ \bibnamefont
  {Dahlhaus}}, \bibinfo {author} {\bibfnamefont {B.~M.}\ \bibnamefont
  {Fregoso}}, \ and\ \bibinfo {author} {\bibfnamefont {J.~E.}\ \bibnamefont
  {Moore}},\ }\bibfield  {title} {\enquote {\bibinfo {title} {Magnetization
  signatures of light-induced quantum hall edge states},}\ }\href {\doibase
  http://arxiv.org/abs/1408.6811} {\bibfield  {journal} {\bibinfo  {journal}
  {arXiv:1408.6811 [cond-mat.mes-hall]}\
  }http://arxiv.org/abs/1408.6811}\BibitemShut {NoStop}%
\bibitem [{\citenamefont {Rudner}\ \emph {et~al.}(2013)\citenamefont {Rudner},
  \citenamefont {Lindner}, \citenamefont {Berg},\ and\ \citenamefont
  {Levin}}]{Rudner2013}%
  \BibitemOpen
  \bibfield  {author} {\bibinfo {author} {\bibfnamefont {M.~S.}\ \bibnamefont
  {Rudner}}, \bibinfo {author} {\bibfnamefont {N.~H.}\ \bibnamefont {Lindner}},
  \bibinfo {author} {\bibfnamefont {E.}~\bibnamefont {Berg}}, \ and\ \bibinfo
  {author} {\bibfnamefont {M.}~\bibnamefont {Levin}},\ }\bibfield  {title}
  {\enquote {\bibinfo {title} {Anomalous edge states and the bulk-edge
  correspondence for periodically driven two-dimensional systems},}\ }\href
  {\doibase 10.1103/PhysRevX.3.031005} {\bibfield  {journal} {\bibinfo
  {journal} {Phys. Rev. X}\ }\textbf {\bibinfo {volume} {3}},\ \bibinfo {pages}
  {031005} (\bibinfo {year} {2013})}\BibitemShut {NoStop}%
\bibitem [{\citenamefont {Ho}\ and\ \citenamefont {Gong}(2014)}]{Ho2014}%
  \BibitemOpen
  \bibfield  {author} {\bibinfo {author} {\bibfnamefont {D.~Y.~H.}\
  \bibnamefont {Ho}}\ and\ \bibinfo {author} {\bibfnamefont {J.}~\bibnamefont
  {Gong}},\ }\bibfield  {title} {\enquote {\bibinfo {title} {Topological
  effects in chiral symmetric driven systems},}\ }\href {\doibase
  10.1103/PhysRevB.90.195419} {\bibfield  {journal} {\bibinfo  {journal} {Phys.
  Rev. B}\ }\textbf {\bibinfo {volume} {90}},\ \bibinfo {pages} {195419}
  (\bibinfo {year} {2014})}\BibitemShut {NoStop}%
\bibitem [{\citenamefont {Yang}()}]{Yang2014}%
  \BibitemOpen
  \bibfield  {author} {\bibinfo {author} {\bibfnamefont {X.}~\bibnamefont
  {Yang}},\ }\bibfield  {title} {\enquote {\bibinfo {title} {Floquet
  topological superfluid and majorana zero modes in two-dimensional
  periodically driven fermi systems},}\ }\href {http://arxiv.org/abs/1410.5035}
  {\bibinfo  {journal} {arXiv:1410.5035 [cond-mat.quant-gas]}\ }\BibitemShut
  {NoStop}%
\bibitem [{\citenamefont {Dehghani}\ \emph {et~al.}(2014)\citenamefont
  {Dehghani}, \citenamefont {Oka},\ and\ \citenamefont {Mitra}}]{Dehghani2014}%
  \BibitemOpen
\bibfield  {journal} {  }\bibfield  {author} {\bibinfo {author} {\bibfnamefont
  {H.}~\bibnamefont {Dehghani}}, \bibinfo {author} {\bibfnamefont
  {T.}~\bibnamefont {Oka}}, \ and\ \bibinfo {author} {\bibfnamefont
  {A.}~\bibnamefont {Mitra}},\ }\bibfield  {title} {\enquote {\bibinfo {title}
  {Dissipative floquet topological systems},}\ }\href
  {http://link.aps.org/doi/10.1103/PhysRevB.90.195429} {\bibfield  {journal}
  {\bibinfo  {journal} {Phys. Rev. B}\ }\textbf {\bibinfo {volume} {90}},\
  \bibinfo {pages} {195429} (\bibinfo {year} {2014})}\BibitemShut {NoStop}%
\bibitem [{\citenamefont {Liu}(2015)}]{Liu2014}%
  \BibitemOpen
  \bibfield  {author} {\bibinfo {author} {\bibfnamefont {D.~E.}\ \bibnamefont
  {Liu}},\ }\bibfield  {title} {\enquote {\bibinfo {title} {Classification of
  the floquet statistical distribution for time-periodic open systems},}\
  }\href {\doibase 10.1103/PhysRevB.91.144301} {\bibfield  {journal} {\bibinfo
  {journal} {Phys. Rev. B}\ }\textbf {\bibinfo {volume} {91}},\ \bibinfo
  {pages} {144301} (\bibinfo {year} {2015})}\BibitemShut {NoStop}%
\bibitem [{\citenamefont {Seetharam}\ \emph {et~al.}()\citenamefont
  {Seetharam}, \citenamefont {Bardyn}, \citenamefont {Lindner}, \citenamefont
  {Rudner},\ and\ \citenamefont {Refael}}]{Seetharam2015}%
  \BibitemOpen
  \bibfield  {author} {\bibinfo {author} {\bibfnamefont {K.~I.}\ \bibnamefont
  {Seetharam}}, \bibinfo {author} {\bibfnamefont {C.-E.}\ \bibnamefont
  {Bardyn}}, \bibinfo {author} {\bibfnamefont {N.~H.}\ \bibnamefont {Lindner}},
  \bibinfo {author} {\bibfnamefont {M.~S.}\ \bibnamefont {Rudner}}, \ and\
  \bibinfo {author} {\bibfnamefont {G.}~\bibnamefont {Refael}},\ }\bibfield
  {title} {\enquote {\bibinfo {title} {Controlled population of floquet-bloch
  states via coupling to bose and fermi baths},}\ }\href
  {http://arxiv.org/abs/1502.02664} {\bibinfo  {journal} {arXiv:1502.02664
  [cond-mat.mes-hall]}\ }\BibitemShut {NoStop}%
\bibitem [{\citenamefont {Iadecola}\ \emph {et~al.}()\citenamefont {Iadecola},
  \citenamefont {Neupert},\ and\ \citenamefont {Chamon}}]{Iadecola2015}%
  \BibitemOpen
\bibfield  {journal} {  }\bibfield  {author} {\bibinfo {author} {\bibfnamefont
  {T.}~\bibnamefont {Iadecola}}, \bibinfo {author} {\bibfnamefont
  {T.}~\bibnamefont {Neupert}}, \ and\ \bibinfo {author} {\bibfnamefont
  {C.}~\bibnamefont {Chamon}},\ }\bibfield  {title} {\enquote {\bibinfo {title}
  {Occupation of topological floquet bands in open systems},}\ }\href
  {http://arxiv.org/abs/1502.05047} {\bibinfo  {journal} {arXiv:1502.05047
  [cond-mat.mes-hall]}\ }\BibitemShut {NoStop}%
\bibitem [{\citenamefont {Dehghani}\ \emph {et~al.}()\citenamefont {Dehghani},
  \citenamefont {Oka},\ and\ \citenamefont {Mitra}}]{Dehghani2015}%
  \BibitemOpen
\bibfield  {journal} {  }\bibfield  {author} {\bibinfo {author} {\bibfnamefont
  {H.}~\bibnamefont {Dehghani}}, \bibinfo {author} {\bibfnamefont
  {T.}~\bibnamefont {Oka}}, \ and\ \bibinfo {author} {\bibfnamefont
  {A.}~\bibnamefont {Mitra}},\ }\bibfield  {title} {\enquote {\bibinfo {title}
  {Out of equilibrium electrons and the hall conductance of a floquet
  topological insulator},}\ }\href {http://arxiv.org/abs/1502.05047} {\bibinfo
  {journal} {arXiv:1412.8469 [cond-mat.mes-hall]}\ }\BibitemShut {NoStop}%
\bibitem [{\citenamefont {Gu}\ \emph {et~al.}(2011)\citenamefont {Gu},
  \citenamefont {Fertig}, \citenamefont {Arovas},\ and\ \citenamefont
  {Auerbach}}]{Gu2011}%
  \BibitemOpen
\bibfield  {journal} {  }\bibfield  {author} {\bibinfo {author} {\bibfnamefont
  {Z.}~\bibnamefont {Gu}}, \bibinfo {author} {\bibfnamefont {H.~A.}\
  \bibnamefont {Fertig}}, \bibinfo {author} {\bibfnamefont {D.~P.}\
  \bibnamefont {Arovas}}, \ and\ \bibinfo {author} {\bibfnamefont
  {A.}~\bibnamefont {Auerbach}},\ }\bibfield  {title} {\enquote {\bibinfo
  {title} {Floquet spectrum and transport through an irradiated graphene
  ribbon},}\ }\href {http://link.aps.org/doi/10.1103/PhysRevLett.107.216601}
  {\bibfield  {journal} {\bibinfo  {journal} {Phys. Rev. Lett.}\ }\textbf
  {\bibinfo {volume} {107}},\ \bibinfo {pages} {216601} (\bibinfo {year}
  {2011})}\BibitemShut {NoStop}%
\bibitem [{\citenamefont {Kundu}\ \emph {et~al.}(2014)\citenamefont {Kundu},
  \citenamefont {Fertig},\ and\ \citenamefont {Seradjeh}}]{Kundu2014}%
  \BibitemOpen
  \bibfield  {author} {\bibinfo {author} {\bibfnamefont {A.}~\bibnamefont
  {Kundu}}, \bibinfo {author} {\bibfnamefont {H.~A.}\ \bibnamefont {Fertig}}, \
  and\ \bibinfo {author} {\bibfnamefont {B.}~\bibnamefont {Seradjeh}},\
  }\bibfield  {title} {\enquote {\bibinfo {title} {Effective theory of floquet
  topological transitions},}\ }\href
  {http://link.aps.org/doi/10.1103/PhysRevLett.113.236803} {\bibfield
  {journal} {\bibinfo  {journal} {Phys. Rev. Lett.}\ }\textbf {\bibinfo
  {volume} {113}},\ \bibinfo {pages} {236803} (\bibinfo {year}
  {2014})}\BibitemShut {NoStop}%
\bibitem [{\citenamefont {Foa~Torres}\ \emph {et~al.}(2014)\citenamefont
  {Foa~Torres}, \citenamefont {Perez-Piskunow}, \citenamefont {Balseiro},\ and\
  \citenamefont {Usaj}}]{Foa2014}%
  \BibitemOpen
  \bibfield  {author} {\bibinfo {author} {\bibfnamefont {L.~E.~F.}\
  \bibnamefont {Foa~Torres}}, \bibinfo {author} {\bibfnamefont {P.~M.}\
  \bibnamefont {Perez-Piskunow}}, \bibinfo {author} {\bibfnamefont {C.~A.}\
  \bibnamefont {Balseiro}}, \ and\ \bibinfo {author} {\bibfnamefont
  {G.}~\bibnamefont {Usaj}},\ }\bibfield  {title} {\enquote {\bibinfo {title}
  {Multiterminal conductance of a floquet topological insulator},}\ }\href
  {http://journals.aps.org/prl/abstract/10.1103/PhysRevLett.113.266801}
  {\bibfield  {journal} {\bibinfo  {journal} {Phys. Rev. Lett.}\ }\textbf
  {\bibinfo {volume} {113}},\ \bibinfo {pages} {266801} (\bibinfo {year}
  {2014})}\BibitemShut {NoStop}%
\bibitem [{\citenamefont {Sambe}(1973)}]{Sambe1973}%
  \BibitemOpen
  \bibfield  {author} {\bibinfo {author} {\bibfnamefont {H.}~\bibnamefont
  {Sambe}},\ }\bibfield  {title} {\enquote {\bibinfo {title} {Steady states and
  quasienergies of a quantum-mechanical system in an oscillating field},}\
  }\href {\doibase 10.1103/PhysRevA.7.2203} {\bibfield  {journal} {\bibinfo
  {journal} {Phys. Rev. A}\ }\textbf {\bibinfo {volume} {7}},\ \bibinfo {pages}
  {2203} (\bibinfo {year} {1973})}\BibitemShut {NoStop}%
\bibitem [{\citenamefont {Shirley}(1965)}]{Shirley1965}%
  \BibitemOpen
  \bibfield  {author} {\bibinfo {author} {\bibfnamefont {J.~H.}\ \bibnamefont
  {Shirley}},\ }\bibfield  {title} {\enquote {\bibinfo {title} {Solution of the
  schr\"odinger equation with a hamiltonian periodic in time},}\ }\href
  {\doibase 10.1103/PhysRev.138.B979} {\bibfield  {journal} {\bibinfo
  {journal} {Phys. Rev.}\ }\textbf {\bibinfo {volume} {138}},\ \bibinfo {pages}
  {B979} (\bibinfo {year} {1965})}\BibitemShut {NoStop}%
\bibitem [{\citenamefont {Grifoni}\ and\ \citenamefont
  {H\"anggi}(1998)}]{Grifoni1998}%
  \BibitemOpen
  \bibfield  {author} {\bibinfo {author} {\bibfnamefont {M.}~\bibnamefont
  {Grifoni}}\ and\ \bibinfo {author} {\bibfnamefont {P.}~\bibnamefont
  {H\"anggi}},\ }\bibfield  {title} {\enquote {\bibinfo {title} {Driven quantum
  tunneling},}\ }\href {\doibase
  http://dx.doi.org/10.1016/S0370-1573(98)00022-2} {\bibfield  {journal}
  {\bibinfo  {journal} {Physics Reports}\ }\textbf {\bibinfo {volume} {304}},\
  \bibinfo {pages} {229 } (\bibinfo {year} {1998})}\BibitemShut {NoStop}%
\bibitem [{\citenamefont {Platero}\ and\ \citenamefont
  {Aguado}(2004)}]{Platero2004}%
  \BibitemOpen
  \bibfield  {author} {\bibinfo {author} {\bibfnamefont {G.}~\bibnamefont
  {Platero}}\ and\ \bibinfo {author} {\bibfnamefont {R.}~\bibnamefont
  {Aguado}},\ }\bibfield  {title} {\enquote {\bibinfo {title} {Photon-assisted
  transport in semiconductor nanostructures},}\ }\href {\doibase
  10.1016/j.physrep.2004.01.004} {\bibfield  {journal} {\bibinfo  {journal}
  {Physics Reports}\ }\textbf {\bibinfo {volume} {395}},\ \bibinfo {pages} {1 }
  (\bibinfo {year} {2004})}\BibitemShut {NoStop}%
\bibitem [{\citenamefont {Kohler}\ \emph {et~al.}(2005)\citenamefont {Kohler},
  \citenamefont {Lehmann},\ and\ \citenamefont {H\"anggi}}]{Kohler2005}%
  \BibitemOpen
  \bibfield  {author} {\bibinfo {author} {\bibfnamefont {S.}~\bibnamefont
  {Kohler}}, \bibinfo {author} {\bibfnamefont {J.}~\bibnamefont {Lehmann}}, \
  and\ \bibinfo {author} {\bibfnamefont {P.}~\bibnamefont {H\"anggi}},\
  }\bibfield  {title} {\enquote {\bibinfo {title} {Driven quantum transport on
  the nanoscale},}\ }\href {\doibase 10.1016/j.physrep.2004.11.002} {\bibfield
  {journal} {\bibinfo  {journal} {Physics Reports}\ }\textbf {\bibinfo {volume}
  {406}},\ \bibinfo {pages} {379} (\bibinfo {year} {2005})}\BibitemShut
  {NoStop}%
\bibitem [{\citenamefont {Eckardt}\ and\ \citenamefont
  {Anisimovas}()}]{Eckardt2015}%
  \BibitemOpen
  \bibfield  {author} {\bibinfo {author} {\bibfnamefont {A.}~\bibnamefont
  {Eckardt}}\ and\ \bibinfo {author} {\bibfnamefont {E.}~\bibnamefont
  {Anisimovas}},\ }\bibfield  {title} {\enquote {\bibinfo {title} {Consistent
  high-frequency approximation for periodically driven quantum systems},}\
  }\href {http://arxiv.org/abs/1502.06477} {\bibinfo  {journal}
  {arXiv:1502.06477 [cond-mat.quant-gas]}\ }\BibitemShut {NoStop}%
\bibitem [{\citenamefont {Wang}\ \emph {et~al.}(2013)\citenamefont {Wang},
  \citenamefont {Steinberg}, \citenamefont {Jarillo-Herrero},\ and\
  \citenamefont {Gedik}}]{Wang2013}%
  \BibitemOpen
\bibfield  {journal} {  }\bibfield  {author} {\bibinfo {author} {\bibfnamefont
  {Y.~H.}\ \bibnamefont {Wang}}, \bibinfo {author} {\bibfnamefont
  {H.}~\bibnamefont {Steinberg}}, \bibinfo {author} {\bibfnamefont
  {P.}~\bibnamefont {Jarillo-Herrero}}, \ and\ \bibinfo {author} {\bibfnamefont
  {N.}~\bibnamefont {Gedik}},\ }\bibfield  {title} {\enquote {\bibinfo {title}
  {Observation of floquet-bloch states on the surface of a topological
  insulator},}\ }\href {\doibase 10.1126/science.1239834} {\bibfield  {journal}
  {\bibinfo  {journal} {Science}\ }\textbf {\bibinfo {volume} {342}},\ \bibinfo
  {pages} {453} (\bibinfo {year} {2013})}\BibitemShut {NoStop}%
\bibitem [{\citenamefont {Tenenbaum~Katan}\ and\ \citenamefont
  {Podolsky}(2013)}]{TenenbaumKatan2013}%
  \BibitemOpen
  \bibfield  {author} {\bibinfo {author} {\bibfnamefont {Y.}~\bibnamefont
  {Tenenbaum~Katan}}\ and\ \bibinfo {author} {\bibfnamefont {D.}~\bibnamefont
  {Podolsky}},\ }\bibfield  {title} {\enquote {\bibinfo {title} {Generation and
  manipulation of localized modes in floquet topological insulators},}\ }\href
  {http://link.aps.org/doi/10.1103/PhysRevB.88.224106} {\bibfield  {journal}
  {\bibinfo  {journal} {Phys. Rev. B}\ }\textbf {\bibinfo {volume} {88}},\
  \bibinfo {pages} {224106} (\bibinfo {year} {2013})}\BibitemShut {NoStop}%
\bibitem [{\citenamefont {G\'omez-Le\'on}\ \emph {et~al.}(2014)\citenamefont
  {G\'omez-Le\'on}, \citenamefont {Delplace},\ and\ \citenamefont
  {Platero}}]{Gomez-Leon2014}%
  \BibitemOpen
  \bibfield  {author} {\bibinfo {author} {\bibfnamefont {A.}~\bibnamefont
  {G\'omez-Le\'on}}, \bibinfo {author} {\bibfnamefont {P.}~\bibnamefont
  {Delplace}}, \ and\ \bibinfo {author} {\bibfnamefont {G.}~\bibnamefont
  {Platero}},\ }\bibfield  {title} {\enquote {\bibinfo {title} {Engineering
  anomalous quantum hall plateaus and antichiral states with ac fields},}\
  }\href {http://link.aps.org/doi/10.1103/PhysRevB.89.205408} {\bibfield
  {journal} {\bibinfo  {journal} {Phys. Rev. B}\ }\textbf {\bibinfo {volume}
  {89}},\ \bibinfo {pages} {205408} (\bibinfo {year} {2014})}\BibitemShut
  {NoStop}%
\bibitem [{\citenamefont {Calvo}\ \emph {et~al.}(2013)\citenamefont {Calvo},
  \citenamefont {Perez-Piskunow}, \citenamefont {Pastawski}, \citenamefont
  {Roche},\ and\ \citenamefont {Foa~Torres}}]{Calvo2013}%
  \BibitemOpen
  \bibfield  {author} {\bibinfo {author} {\bibfnamefont {H.~L.}\ \bibnamefont
  {Calvo}}, \bibinfo {author} {\bibfnamefont {P.~M.}\ \bibnamefont
  {Perez-Piskunow}}, \bibinfo {author} {\bibfnamefont {H.~M.}\ \bibnamefont
  {Pastawski}}, \bibinfo {author} {\bibfnamefont {S.}~\bibnamefont {Roche}}, \
  and\ \bibinfo {author} {\bibfnamefont {L.~E.~F.}\ \bibnamefont
  {Foa~Torres}},\ }\bibfield  {title} {\enquote {\bibinfo {title}
  {Non-perturbative effects of laser illumination on the electrical properties
  of graphene nanoribbons},}\ }\href
  {http://stacks.iop.org/0953-8984/25/i=14/a=144202} {\bibfield  {journal}
  {\bibinfo  {journal} {Journal of Physics: Condensed Matter}\ }\textbf
  {\bibinfo {volume} {25}},\ \bibinfo {pages} {144202} (\bibinfo {year}
  {2013})}\BibitemShut {NoStop}%
\bibitem [{\citenamefont {Koghee}\ \emph {et~al.}(2012)\citenamefont {Koghee},
  \citenamefont {Lim}, \citenamefont {Goerbig},\ and\ \citenamefont
  {Smith}}]{Koghee2012}%
  \BibitemOpen
  \bibfield  {author} {\bibinfo {author} {\bibfnamefont {S.}~\bibnamefont
  {Koghee}}, \bibinfo {author} {\bibfnamefont {L.-K.}\ \bibnamefont {Lim}},
  \bibinfo {author} {\bibfnamefont {M.~O.}\ \bibnamefont {Goerbig}}, \ and\
  \bibinfo {author} {\bibfnamefont {C.~M.}\ \bibnamefont {Smith}},\ }\bibfield
  {title} {\enquote {\bibinfo {title} {Merging and alignment of dirac points in
  a shaken honeycomb optical lattice},}\ }\href {\doibase
  10.1103/PhysRevA.85.023637} {\bibfield  {journal} {\bibinfo  {journal} {Phys.
  Rev. A}\ }\textbf {\bibinfo {volume} {85}},\ \bibinfo {pages} {023637}
  (\bibinfo {year} {2012})}\BibitemShut {NoStop}%
\bibitem [{\citenamefont {Delplace}\ \emph {et~al.}(2013)\citenamefont
  {Delplace}, \citenamefont {G\'{o}mez-Le\'{o}n},\ and\ \citenamefont
  {Platero}}]{Delplace2013}%
  \BibitemOpen
  \bibfield  {author} {\bibinfo {author} {\bibfnamefont {P.}~\bibnamefont
  {Delplace}}, \bibinfo {author} {\bibfnamefont {A.}~\bibnamefont
  {G\'{o}mez-Le\'{o}n}}, \ and\ \bibinfo {author} {\bibfnamefont
  {G.}~\bibnamefont {Platero}},\ }\bibfield  {title} {\enquote {\bibinfo
  {title} {{Merging of Dirac points and Floquet topological transitions in
  ac-driven graphene}},}\ }\href {\doibase 10.1103/PhysRevB.88.245422}
  {\bibfield  {journal} {\bibinfo  {journal} {Physical Review B}\ }\textbf
  {\bibinfo {volume} {88}},\ \bibinfo {pages} {245422} (\bibinfo {year}
  {2013})}\BibitemShut {NoStop}%
\bibitem [{\citenamefont {Savel'ev}\ and\ \citenamefont
  {Alexandrov}(2011)}]{Savelev2011}%
  \BibitemOpen
  \bibfield  {author} {\bibinfo {author} {\bibfnamefont {S.~E.}\ \bibnamefont
  {Savel'ev}}\ and\ \bibinfo {author} {\bibfnamefont {A.~S.}\ \bibnamefont
  {Alexandrov}},\ }\bibfield  {title} {\enquote {\bibinfo {title} {Massless
  dirac fermions in a laser field as a counterpart of graphene
  superlattices},}\ }\href {http://link.aps.org/doi/10.1103/PhysRevB.84.035428}
  {\bibfield  {journal} {\bibinfo  {journal} {Phys. Rev. B}\ }\textbf {\bibinfo
  {volume} {84}},\ \bibinfo {pages} {035428} (\bibinfo {year}
  {2011})}\BibitemShut {NoStop}%
\bibitem [{\citenamefont {Baum}\ \emph {et~al.}(2015)\citenamefont {Baum},
  \citenamefont {Posske}, \citenamefont {Fulga}, \citenamefont {Trauzettel},\
  and\ \citenamefont {Stern}}]{Baum2015}%
  \BibitemOpen
  \bibfield  {author} {\bibinfo {author} {\bibfnamefont {Y.}~\bibnamefont
  {Baum}}, \bibinfo {author} {\bibfnamefont {T.}~\bibnamefont {Posske}},
  \bibinfo {author} {\bibfnamefont {I.~C.}\ \bibnamefont {Fulga}}, \bibinfo
  {author} {\bibfnamefont {B.}~\bibnamefont {Trauzettel}}, \ and\ \bibinfo
  {author} {\bibfnamefont {A.}~\bibnamefont {Stern}},\ }\bibfield  {title}
  {\enquote {\bibinfo {title} {Coexisting edge states and gapless bulk in
  topological states of matter},}\ }\href {\doibase
  10.1103/PhysRevLett.114.136801} {\bibfield  {journal} {\bibinfo  {journal}
  {Phys. Rev. Lett.}\ }\textbf {\bibinfo {volume} {114}},\ \bibinfo {pages}
  {136801} (\bibinfo {year} {2015})}\BibitemShut {NoStop}%
\bibitem [{\citenamefont {Thouless}\ \emph {et~al.}(1982)\citenamefont
  {Thouless}, \citenamefont {Kohmoto}, \citenamefont {Nightingale},\ and\
  \citenamefont {den Nijs}}]{Thouless1982}%
  \BibitemOpen
  \bibfield  {author} {\bibinfo {author} {\bibfnamefont {D.~J.}\ \bibnamefont
  {Thouless}}, \bibinfo {author} {\bibfnamefont {M.}~\bibnamefont {Kohmoto}},
  \bibinfo {author} {\bibfnamefont {M.~P.}\ \bibnamefont {Nightingale}}, \ and\
  \bibinfo {author} {\bibfnamefont {M.}~\bibnamefont {den Nijs}},\ }\bibfield
  {title} {\enquote {\bibinfo {title} {Quantized hall conductance in a
  two-dimensional periodic potential},}\ }\href
  {http://link.aps.org/doi/10.1103/PhysRevLett.49.405} {\bibfield  {journal}
  {\bibinfo  {journal} {Phys. Rev. Lett.}\ }\textbf {\bibinfo {volume} {49}},\
  \bibinfo {pages} {405} (\bibinfo {year} {1982})}\BibitemShut {NoStop}%
\bibitem [{\citenamefont {Aidelsburger}\ \emph {et~al.}(2014)\citenamefont
  {Aidelsburger}, \citenamefont {Lohse}, \citenamefont {Schweizer},
  \citenamefont {Atala}, \citenamefont {Barreiro}, \citenamefont
  {Nascimb\`{e}ne}, \citenamefont {Cooper}, \citenamefont {Bloch},\ and\
  \citenamefont {Goldman}}]{Aidelsburger2015}%
  \BibitemOpen
  \bibfield  {author} {\bibinfo {author} {\bibfnamefont {M.}~\bibnamefont
  {Aidelsburger}}, \bibinfo {author} {\bibfnamefont {M.}~\bibnamefont {Lohse}},
  \bibinfo {author} {\bibfnamefont {C.}~\bibnamefont {Schweizer}}, \bibinfo
  {author} {\bibfnamefont {M.}~\bibnamefont {Atala}}, \bibinfo {author}
  {\bibfnamefont {J.~T.}\ \bibnamefont {Barreiro}}, \bibinfo {author}
  {\bibfnamefont {S.}~\bibnamefont {Nascimb\`{e}ne}}, \bibinfo {author}
  {\bibfnamefont {N.~R.}\ \bibnamefont {Cooper}}, \bibinfo {author}
  {\bibfnamefont {I.}~\bibnamefont {Bloch}}, \ and\ \bibinfo {author}
  {\bibfnamefont {N.}~\bibnamefont {Goldman}},\ }\bibfield  {title} {\enquote
  {\bibinfo {title} {{Measuring the Chern number of Hofstadter bands with
  ultracold bosonic atoms}},}\ }\href {\doibase 10.1038/NPHYS3171} {\bibfield
  {journal} {\bibinfo  {journal} {Nat. Phys.}\ }\textbf {\bibinfo {volume}
  {11}},\ \bibinfo {pages} {1} (\bibinfo {year} {2014})},\ \Eprint
  {http://arxiv.org/abs/1407.4205} {arXiv:1407.4205} \BibitemShut {NoStop}%
\bibitem [{Note1()}]{Note1}%
  \BibitemOpen
  \bibinfo {note} {\label {note}To keep the expressions simple we have
  neglected corrections of order $\eta ^2$ for the $z$ components of $\protect
  \bm {h}$. These corrections come from a renormalization of the Floquet
  replicas and do not modify the contribution to the Chern number.}\BibitemShut
  {Stop}%
\bibitem [{\citenamefont {Gomez-Leon}\ and\ \citenamefont
  {Platero}(2013)}]{Gomez-Leon2013}%
  \BibitemOpen
  \bibfield  {author} {\bibinfo {author} {\bibfnamefont {A.}~\bibnamefont
  {Gomez-Leon}}\ and\ \bibinfo {author} {\bibfnamefont {G.}~\bibnamefont
  {Platero}},\ }\bibfield  {title} {\enquote {\bibinfo {title} {Floquet-bloch
  theory and topology in periodically driven lattices},}\ }\href
  {http://link.aps.org/doi/10.1103/PhysRevLett.110.200403} {\bibfield
  {journal} {\bibinfo  {journal} {Phys. Rev. Lett.}\ }\textbf {\bibinfo
  {volume} {110}},\ \bibinfo {pages} {200403} (\bibinfo {year}
  {2013})}\BibitemShut {NoStop}%
\end{thebibliography}%

\end{document}